\documentclass[11pt,a4paper]{article}
\usepackage{jheppub}
\usepackage[T1]{fontenc}
\usepackage{lmodern}
\usepackage{mathrsfs}
\usepackage{dsfont}
\usepackage{mathtools,bm}
\usepackage{booktabs,array,longtable}
\usepackage{graphicx,xcolor,tikz}
\usetikzlibrary{arrows.meta,positioning,calc}
\usepackage{microtype}
\numberwithin{equation}{section}
\newcommand{\one}{\mathbf1}
\newcommand{\Is}{\mathrm{Is}}
\newcommand{\TIM}{\mathrm{TIM}}
\newcommand{\qdim}{\operatorname{qdim}}

\newcommand{\Res}{\operatorname{Res}}
\newcommand{\Rep}{\operatorname{Rep}}
\newcommand{\tr}{\operatorname{Tr}}
\newcommand{\ket}[1]{\left|#1\right\rangle}
\newcommand{\bra}[1]{\langle#1|}
\newcommand{\ish}[1]{|#1\rangle\!\rangle}
\newcommand{\cI}{\mathcal I}
\newcommand{\cL}{\mathcal L}
\newcommand{\fA}{\mathfrak a_A}
\newcommand{\fphi}{\varphi}

\newcommand{\Ttot}{T_{\mathrm{tot}}}

\preprint{USTC-ICTS/PCFT-26-84}
\title{Rational conformal interfaces between
the Ising and tricritical Ising models  from a tetrality web}
\author[A]{Ant\'onio Antunes,}
\author[B,C]{Jin Chen}
\author[D]{and Junchen Rong}
\affiliation[A]{Laboratoire de Physique de l'\'Ecole Normale Sup\'erieure, Universit\'e  PSL, CNRS, Sorbonne Universit\'e, Universit\'e  Paris Cit\'e, 24 rue Lhomond, F-75005 Paris, France}
\affiliation[B]{Department of Physics, Xiamen University, Xiamen 361005, China}
\affiliation[C]{Peng Huanwu Center for Fundamental Theory, Hefei, Anhui 230026, China}
\affiliation[D]{CPHT, CNRS, \'Ecole Polytechnique, Institut Polytechnique de Paris, Palaiseau, France}
\emailAdd{junchen.rong@polytechnique.edu, antonio.antunes@ens.fr, zenofox@gmail.com}
\abstract{
We construct rational conformal interfaces between the Ising
and tricritical Ising (TIM) models by using different chiral symmetries to prepare their folded boundary states. We use a \emph{moving up} procedure which
extends the product theory to an auxiliary coset with an
$S_3$ automorphism group, where permutations give twisted
Cardy states. We complement this with a \emph{moving down} approach which reduces to the $S_3$ invariant algebra.
Overall, these procedures uncover a \emph{tetrality} web relating four different theories. Pulling back ordinary Cardy states of the smallest theory in the web to the physical tensor product system gives eighteen
factorized interfaces and six transmitting interfaces.
The latter agree with four transposition and two 3-cycle
twisted boundaries in the largest theory. They include Gaiotto's
RG wall and the interface obtained by fusing an Ising
duality line into it. The six interfaces carry the regular
action of the TIM fusion ring and have
the same energy transmission coefficient, $\frac{35}{48}$. Their boundary entropies,
bulk correlations and operator spectra distinguish them.
We calculate these observables and the complete annulus
amplitudes between the twenty four boundaries.
}
\begin{document}
\maketitle
\section{Introduction}\label{sec:intro}

A conformal interface is a co-dimension one extended object that connects two conformal field theories (CFTs) and preserves the diagonal conformal symmetry of the two bulk systems along the defect~\cite{Bachas:2001vj,Quella:2002ct}. The correlation functions in the presence of the interface relate the operator content of the two bulk theories. An important case are the correlators of the two bulk stress tensors which measure the energy transmission and reflection rates across the interface~\cite{Quella:2006de}. Moreover,  interfaces deserve to be studied on their own right: they support their own interface operators, whose dimensions and operator product expansion (OPE) coefficients govern the local excitations and deformations that can trigger interface renormalization group (RG) flows solely along the interface itself~\cite{Kormos:2009sk,Rong:2026xxg}. A particularly interesting example is an RG interface, associated with a bulk RG flow, joining a bulk ultraviolet (UV) theory and its infrared (IR) fixed point and encoding non-perturbative information of the UV--IR operator map~\cite{Brunner:2007ur,Douglas:2010ic,Gaiotto:2012np,Cogburn:2023xzw,Zhang:2026gqp}. However, finding these interfaces is a difficult task even for rational CFTs in two dimensional spacetime, as the interface itself may be intrinsically irrational even when the  bulk theories are rational. For example, so far, the Ising model is the only case among 2d unitary minimal models for which its conformal interfaces/defects (an interface between two copies of the same theory is usually referred to as a defect) have been fully classified~\cite{Oshikawa:1996dj, Kormos:2009sk,Bachas:2013ora}. Employing the folding trick, one obtains a doubled
Ising theory which can be described by a special point in the orbifold branch of a free boson. A boundary state of the tensor product theory then converts to an interface in between the two Ising models. The boson admits Dirichlet and Neumann boundary families, either of which is parameterized by a real modulus parameter. When the modulus is irrational, the interfaces contains infinitely many primaries with respect to the diagonal Virasoro algebra and thus becomes irrational. 

Generalized symmetries describe symmetry operations through topological defects and include higher form symmetries acting on extended operators~\cite{Gaiotto:2014kfa}. In two dimensions, topological defect lines can fuse into sums of lines, realizing noninvertible symmetries. A finite collection closed under fusion, together with its junction data, forms a fusion category~\cite{Frohlich:2006ch,Bhardwaj:2017xup,Chang:2018iay, Thorngren:2021yso,Shao:2023gho}. When preserved by a bulk perturbation, these lines survive along the RG flow and constrain the infrared theory. Matching their fusion, junction and anomaly data between the ultraviolet and infrared fixed points has led to proposed massless flows among minimal models and coset CFTs~\cite{Nakayama:2024msv,Ambrosino:2025yug,Gaberdiel:2026sfg,Ambrosino:2026umb,Benedetti:2026drn,Fukusumi:2026pkr}. The same symmetry matching constrains the lines transmitted across an RG interface and the allowed couplings between its two sides~\cite{Gaiotto:2012np,Chen:2026byx,Fukusumi:2026pkr,Fukusumi:2026tii}. For broader discussions of noninvertible symmetries in two dimensions, see~\cite{Fuchs:2001qc,Fuchs:2002cm,Frohlich:2004ef,Fuchs:2012dt,Komargodski:2020mxz,Thorngren:2019iar,Kaidi:2022cpf,Chang:2022hud,Kaidi:2023maf,Bhardwaj:2023ayw,Bhardwaj:2023kri,Chen:2023qnv,Luo:2023ive,Cordova:2024iti,Chen:2025qub,Antinucci:2025fjp,Chen:2026wyn,Chen:2026cqz}.

In this paper, we will extend this picture to conformal interfaces between the Ising and tri-critical Ising models (TIM). We concentrate on rational interfaces, by which we mean that the folded boundary state preserves a rational chiral algebra containing the total stress tensor such that the spectrum of its self-annulus partition function is organized into finitely many irreducible representations of the algebra, given by a sum of the corresponding characters with nonnegative integer multiplicities~\cite{Fuchs:2002cm}. This gives a way to retain analytic control over the interface spectrum while allowing energy to pass between the two bulk theories. It is worth noting that, in the folded picture, the ordinary Cardy states for the full tensor product algebra of the two bulk theories only give independent boundaries on the two sides. The corresponding interface is hence totally reflective. In contrast, we will seek non-factorizable conformal interfaces with non-zero energy transmission rates between the two bulk theories. The RG interface constructed by Gaiotto is an example of precisely this situation~\cite{Gaiotto:2012np,Antunes:2026kqb}. Consider the folded theory $A=\mathrm{Ising}\times\overline{\mathrm{TIM}}$, where the bar denotes the reversed orientation of the TIM. The associated chiral algebra is a tensor product of the two Virasoro algebras and its central charge is $c_A=\frac12+\frac{7}{10}=\frac65$. It can be conformally embedded into an enlarged chiral algebra, denoted by $B$, via orbifolding of the diagonal $\mathbb Z_2$ Verlinde lines of Ising and TIM,
\begin{align}
 A
 %=\frac{\widehat{\mathfrak{su}}(2)_1^{(1)}
 %\times\widehat{\mathfrak{su}}(2)_1^{(2)}}
 %{\widehat{\mathfrak{su}}(2)_2}\times
 %\frac{\widehat{\mathfrak{su}}(2)_2
 %\times\widehat{\mathfrak{su}}(2)_1^{(3)}}
 %{\widehat{\mathfrak{su}}(2)_3}
\quad\xhookrightarrow{\ \ \ \ }\quad
 B=\frac{\widehat{\mathfrak{su}}(2)_1^{(1)}
 \times\widehat{\mathfrak{su}}(2)_1^{(2)}
 \times\widehat{\mathfrak{su}}(2)_1^{(3)}}
 {\widehat{\mathfrak{su}}(2)_3}\,,
 \label{eq:A_to_B}
\end{align}
where the three superscripts label the numerator factors and the denominator is their diagonal affine algebra. The chiral algebra $B$ admits an $S_3$ automorphism group by permutations of its three numerator factors. In~\cite{Gaiotto:2012np}, it is shown that an automorphism twisted Cardy state defined by the transposition $\sigma_{23}$ of the second and third $\widehat{\mathfrak{su}}(2)_1^{}$ factors plays a role of mixing the Ising and TIM stress tensors. So, its pullback to $A$, through the topological interface associated with the extension~\eqref{eq:A_to_B}, results in a non-factorizable physical interface that transmits energy between Ising and TIM. We refer to this procedure as \emph{moving up}: enlarge the chiral algebra from the physical product theory $A$ to an auxiliary theory $B$ by possible conformal embeddings, construct non-trivial twisted boundary states in $B$, and pull back the boundary states to the physical theory $A$. The chiral operators attached to transmitted symmetry lines, called phantom currents, give a physical description of this extension. Currents of spin one can generate conformal manifolds of interfaces~\cite{Antinucci:2025uvj,Zhang:2026gqp}, while currents of spin two mix with the stress tensors and determine energy transmission~\cite{Furuta:2025ahl}. One of the authors recently developed the construction systematically using these currents and transmitted noninvertible symmetries~\cite{Chen:2026byx}. It constructs generic conformal interfaces beyond RG walls and applies to unitary as well as nonunitary rational CFTs, where varying the interface modulus can lead to logarithmic interfaces~\cite{Chen:2026nac}.

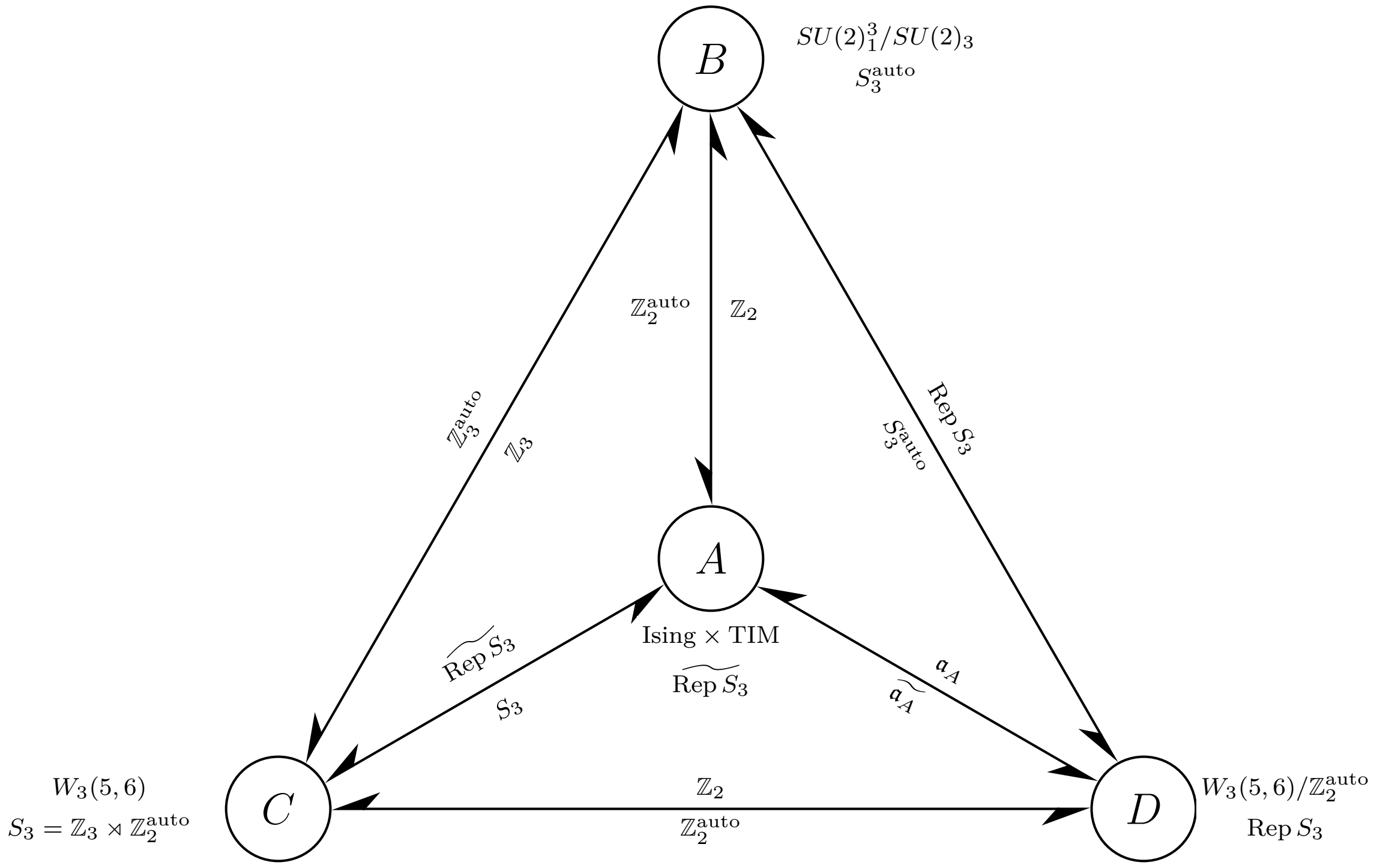
\begin{figure}[t]
\centering
\resizebox{\linewidth}{!}{%
\begin{tikzpicture}[
  scale=1.0,
  theory/.style={circle,draw,thick,fill=white,minimum size=1.20cm,
    inner sep=1pt,font=\Large},
  edge/.style={thick,
    {Stealth[harpoon,swap,length=5.2mm,width=3.8mm]}-{Stealth[harpoon,swap,length=5.2mm,width=3.8mm]}},
  lab/.style={font=\scriptsize,fill=white,inner sep=1.5pt,align=center}
]
  % Geometry from duality_web_note.tex, with the theories relocated:
  % old top A -> B, old centre D -> A, old right B -> D; C unchanged.
  \node[theory] (B) at (0,5.75) {$B$};
  \node[theory] (C) at (-4.98,-2.88) {$C$};
  \node[theory] (D) at (4.98,-2.88) {$D$};
  \node[theory] (A) at (0,0) {$A$};

  \draw[edge] (B)--(D);
  \draw[edge] (B)--(C);
  \draw[edge] (D)--(C);
  \draw[edge] (B)--(A);
  \draw[edge] (D)--(A);
  \draw[edge] (C)--(A);

  % D -> C above the horizontal edge, C -> D below it.
  \node[lab] at (0,-2.64) {$\mathbb Z_2$};
  \node[lab] at (0,-3.12) {$\mathbb Z_2^{\mathrm{auto}}$};
  % C -> B on the inner side, B -> C on the outer side.
  \node[lab,rotate=60] at (-2.21,1.28) {$\mathbb Z_3$};
  \node[lab,rotate=60] at (-2.77,1.60)
    {$\mathbb Z_3^{\mathrm{auto}}$};
  % B -> D on the inner side, D -> B on the outer side.
  \node[lab,rotate=-60] at (2.21,1.28) {$S_3^{\mathrm{auto}}$};
  \node[lab,rotate=-60] at (2.77,1.60) {$\Rep S_3$};
  % C -> A below the diagonal, A -> C above it.
  \node[lab,rotate=30,inner sep=1pt] at (-2.33,-1.72) {$S_3$};
  \node[lab,rotate=30,inner sep=1pt] at (-2.69,-1.09)
    {$\widetilde{\Rep S_3}$};
  % A -> D below the diagonal, D -> A above it.
  \node[lab,rotate=-30] at (2.24,-1.58) {$\widetilde{\fA}$};
  \node[lab,rotate=-30] at (2.74,-1.30) {$\fA$};
  % B -> A on the left of the vertical edge, A -> B on its right.
  \node[lab,anchor=east] at (-0.16,2.86) {$\mathbb Z_2^{\mathrm{auto}}$};
  \node[lab,anchor=west] at (0.16,2.86) {$\mathbb Z_2$};

  \node[lab,anchor=west] at (0.92,5.75)
    {$\begin{gathered}SU(2)_1^3/SU(2)_3\\
      S_3^{\mathrm{auto}}\end{gathered}$};
  \node[lab,anchor=west] at (5.58,-2.88)
    {$\begin{gathered}W_3(5,6)/\mathbb Z_2^{\mathrm{auto}}\\
      \Rep S_3\end{gathered}$};
  \node[lab,anchor=east] at (-5.92,-2.88)
    {$\begin{gathered}W_3(5,6)\\
      S_3=\mathbb Z_3\rtimes\mathbb Z_2^{\mathrm{auto}}
      \end{gathered}$};
  \node[lab] at (0,-1.18)
    {$\begin{gathered}\mathrm{Ising}\times\mathrm{TIM}\\
      \widetilde{\Rep S_3}\end{gathered}$};
\end{tikzpicture}%
}
\caption{The four theories and the gaugings connecting them. Each gauging label belongs to the harpoon on the corresponding side of the edge. The superscript $\mathrm{auto}$ denotes a symmetry acting by an automorphism of the chiral algebra. In $D$, let $\one$ be the vacuum and $z,J$ be the modules with lowest weights three and two, respectively. Extending by $\one\oplus z$, $\fA=\one\oplus J$ and $\one\oplus z\oplus2J$ gives $C$, $A$ and $B$, respectively. The last extension is the regular algebra of $\Rep S_3$. On $A$, gauging the dual algebra $\widetilde{\fA}$ returns to $D$, while $\widetilde{\Rep S_3}$ reverses the $S_3$ gauging from $C$ to $A$.}
\label{fig:web}
\end{figure}

On the other hand, in this paper, we also develop the complementary procedure of \emph{moving down}. We look for a rational chiral algebra $D$ contained in $A$ and sharing its stress tensor. Since the total Virasoro algebra at $c=\frac65$ is not rational, the algebra $D$ in this construction must contain additional currents beyond the stress tensor. An ordinary Cardy state of $D$ glues fewer currents than a product Cardy state of $A$, leaving room for the Ising and TIM energy densities to mix at the boundary~\cite{Quella:2002ct}. Similarly, the conformal embedding $D\subset A$ defines a topological interface through which we can transport these Cardy states back to $A$. The pullbacks are usually a direct sum of simple branes, whose elementary components give the physical interfaces. For Ising and TIM, the moving up and moving down constructions are connected by a tetrahedral duality web depicted in Figure~\ref{fig:web}. We refer to this as the \emph{tetrality} web. More generally, gauging organizes CFTs into networks of related theories~\cite{Gaiotto:2020iye,Diatlyk:2023fwf}. It produces a dual quantum symmetry whose gauging recovers the original theory~\cite{Bhardwaj:2017xup,Chen:2026cqz}. Each node can be obtained from gauging a subgroup $H$ of the $S_3$ automorphisms of $B$ -- which has the maximal chiral algebra in the tetrality. Writing $B^H$ for the corresponding orbifolds and $\langle \sigma_{ij\dots}\rangle$ for the subgroup generated by the cyclic permutations $(ij\dots)$, we have
\begin{align}
 A=B^{\langle\sigma_{12}\rangle},\qquad
 C=B^{\langle\sigma_{123}\rangle}=W_3(5,6),\qquad
 D=B^{S_3}=C^{\mathsf{cc}}\,,
 \label{eq:introweb}
\end{align}
where $W_3(5,6)$ is the second $W_3$ minimal model with central charge $\frac{6}{5}$, and $\mathsf{cc}$ denotes the charge conjugation which reverses its spin three current. The $A$, $B$, $C$ and $D$ theories have $18$, $8$, $20$ and $24$ irreducible modules, respectively. The algebra $D$ is the smallest one in this web. It is strongly generated by currents of spin six, eight and ten and has no nontrivial conformal automorphisms. How much further can the tetrality web be extended? The OPEs of $D$ provide a strong restriction. Our exact calculations, developed with AI assistance, show that a proper strongly rational conformal subalgebra $U$ between $\mathrm{Vir}_{\frac65}$ and $D$ must have its first extra vacuum primary at weight at least twenty five. Here $\mathrm{Vir}_{\frac65}$ is generated by the sum of the stress tensor of Ising and TIM, which is also the stress tensor of $U$. We therefore conjecture that no non-trivial $U$ exists and that the tetrality web of Figure~\ref{fig:web} is complete.

Since the theory $D$ has $24$ modules, in the moving down approach we can start with the corresponding $24$ Cardy states of $D$ and pull them back into boundary states of theory $A$. This gives $18$ tensor products of Ising and TIM Cardy states and $6$ genuine non-factorizable interfaces. The product states are totally reflective, whereas the other $6$ have nonzero transmission rates. On the other hand, the moving up approach from $B$ reproduces the same answer: Different Cardy labels for $\sigma_{23}$-twisted Cardy states give $4$ non-factorizable interfaces, and a $\sigma_{123}$-twist gives the remaining $2$. Non-invertible symmetries can further explain how these six interfaces are related. A Verlinde line in TIM can be brought to the interface and fused into it. In a suitable basis of the $6$ boundaries in the folded picture, they exactly furnish the regular NIM representation of the fusion category of TIM, which can be factorized as the product of the Ising and Fibonacci fusion rings. All $6$ interfaces have energy transmission $\mathsf T=\frac{35}{48}$, but their boundary entropies, correlations with bulk fields and interface operator spectra differ from each other. We calculate these quantities and obtain every self and mixed annulus among the twenty four boundaries as a finite sum of $D$ characters.

We organize the paper as follows: Section~\ref{sec:methods} introduces the general moving up and moving down constructions. Section~\ref{sec:common} describes the tetrality web and the currents preserved by its boundary states.
Section~\ref{sec:construction} constructs the physical interfaces from $D$ Cardy states and from twisted Cardy states in $B$. In Section~\ref{sec:spectra} we study the action of topological lines, the
interface spectra and the physical observables. Section~\ref{sec:discussion} discusses applications and further directions. Details of the OPE calculations, modular data, and character formulas are collected in the appendices.

\paragraph{Note added:}
We would like to draw the reader's attention to the independent work of Bufan Zheng ``Classification of Conformal Ising/Tricritical Ising Interfaces Preserving $\mathcal{W}_3^{\mathbb{Z}_2}$ Symmetry '', which investigates closely related questions and is being submitted to arXiv concurrently with our manuscript. We thank the author for coordinating the submissions.

\section{Building the web of vertex operator algebras}\label{sec:methods}

\subsection{Why build a web?}
When we fold the tricritical Ising model (TIM) across the interface, it becomes a boundary of the diagonal product theory $A$, whose vacuum chiral algebra and stress tensor are 
\[
A=\mathrm{Ising}\otimes\overline{\mathrm{TIM}},\qquad
 \Ttot=T_{\rm I}+\overline{T}_{\rm TIM},\qquad c=\tfrac12+\tfrac7{10}=\tfrac65 .
\]
A conformal boundary glues $\Ttot$ across its two chiralities. 
Which other rational symmetries could the interface preserve? We search for such symmetries by the following approach. 
Start at the folded theory $A$ and repeatedly take a local extension or a rational conformal subalgebra. 
The connected collection is the \emph{web}. We take its nodes to be local, 
positive strongly rational vertex operator algebras (VOAs), up to isomorphism. ``Strongly rational'' means that there are finitely many inequivalent irreducible sectors and every admissible representation decomposes into a direct sum of them.
``Positive'' means that the weights of all primaries are positive.
An edge in the web records an actual inclusion with the same stress tensor. 
At every new node, search in \emph{both} directions: each node can have multiple parents and multiple children.

A \emph{bottom node} has no proper admissible rational conformal subalgebra (except for the Virasoro algebra itself). 
If a boundary preserves a VOA $U$, it also preserves every $R\subset U$, including Virasoro.
We can then use the bottom node to build rational boundaries of the folded theory. 
If the bottom node is rational at the total central charge, then it has finitely many Cardy states. 
We can use the arrows of the web to pull these Cardy states back to conformal interfaces of the folded theory $A$.
For two generic CFTs, the web may contain more than one bottom node and it is necessary to identify duplicated interfaces from different bottom nodes. Fortunately, in our $c=\frac{6}{5}$ case, this does not happen.

The method above allows us to perform an extensive search for rational conformal interfaces between the Ising and TIM. 
However, it is not necessarily exhaustive: there could be different disconnected webs. 
Below, we will also see that the down move is performed by an algorithm that in practice is not exhaustive either as it involves an unbounded search.

\subsection{Moving up: extend the vacuum algebra}
For a rational algebra $V$, write the proposed extended vacuum as a sum of irreducible $V$-modules, $W|_V=\one\oplus\bigoplus_{i\ne0}m_i V_i$, where $\one$ is the $V$ vacuum module. 
Keep only integer lowest weights~\footnote{One may wish to include half-integer weights. However, by Proposition~2.4 of H{\"o}hn and M{\"o}ller~\cite{Hohn:2023auw},
the parity-even subalgebra \(V_{\bar 0}\) of any strongly rational vertex operator superalgebra \(V\) is itself strongly rational.}, since $W$ is an ordinary local vacuum algebra with the same $L_0$. 
Then find the OPE multiplications in this sum that are: (1) unital: the original identity remains identity, (2) associative, and (3) local~\cite{Huang:2014ixa}.
We also need to check that the resulting algebra qualifies as a node of our web. 
In particular, \(W\) must itself be positive and strongly rational.
If every added module is invertible under fusion, this is a \emph{simple-current extension}; otherwise it is a \emph{non-simple-current extension}. Both occur in our $c=\frac{6}{5}$ example. In $A$, the product $j=I_{\frac{1}{2}}\otimes T_{\frac{3}{2}}$ of the indicated Ising and TIM representations has weight two and $j\boxtimes_A j=\one$. Its local extension is $B=A\oplus j$.
By contrast, the inclusion $D\subset A$ below adds a weight-two module of quantum dimension two, so it is not a simple-current extension.

\subsection{Moving down: test the first extra current}\label{downmove}
A restriction selects fields whose OPE closes and retains the stress tensor. 
Taking the invariant fields $V^G$ of a finite automorphism group gives explicit examples.
But automorphisms need not reveal every restriction: Theory $A$ has a  trivial automorphism group and still contains $D$. To search without guessing a group, we can ask how low the spin of the first extra current of a subalgebra $U\subseteq V$ can be:
\[
 h_*(U)=\min\{h>0:V_h\subset U\},
\]
 for a \emph{total} Virasoro primary $V_h$.
 We can then follow the algorithm below:

 \medskip

\textbf{1. Find all candidates at weight $h$.} In the vacuum space of $V$, remove null states and compute $\ker L_1\cap\ker L_2$ at weight $h$. These conditions imply $L_n v=0$ for every $n>0$. For a basis $p_1,\ldots,p_n$ of this entire primary space, write $v=\sum_i x_i p_i$, with arbitrary complex coefficients. This includes mixtures of different symmetry sectors.

\textbf{2. Ask what the candidate forces into $U$.} By definition, every operator in the OPE $v(z)v(0)$ must lie in $U$. At each weight $0<k<h$, descendants of the identity are allowed, but the projection onto lower primaries must vanish if $v$ is the lightest primary. Each component gives a homogeneous quadratic equation in the $x_i$: the OPE is linear in each of its two inputs. These are necessary conditions on \emph{every} vector in the space of $V_h$'s, even when there is more than one current of spin $h$.

\textbf{Example: could a subalgebra of $B$ start at weight four?}
There are two weight-four primaries in $B$ after removing null states. Write $v=x_0p_0+x_1p_1$. The coefficient of $z^{-6}$ in $v(z)v(0)$ has weight two. Subtract its component along $\Ttot$, which already belongs to $U$. The remainder is a weight-two primary in $U$ and so must vanish. In one choice of basis its two components give
\[
 x_0x_1=0,\qquad 27x_0^2-16x_1^2=0.
\]
The first equation sets one coefficient to zero and the second sets the other to zero. A subalgebra may \emph{contain} weight-four primaries, but none can be its \emph{lightest}: each nonzero choice brings in a weight-two current. 

\textbf{3. Exclude the weight or follow the surviving OPEs.} At weight four we solved the two equations by inspection. For larger primary spaces it is easier to find polynomial conditions that force a power of every coordinate $x_i$ to vanish. In Appendix \ref{app:opeconstraints}, we give another illustrative example at weight 21.
If nonzero candidates survive, further OPEs (which lead to higher degree polynomial equations) may force a known VOA or rule out the proposed first weight.  Passing the quadratic test alone does not establish rationality.

\subsection{Results for the Ising--TIM search}
The largest theory in the known web is the coset $B=SU(2)_1^3/SU(2)_3$. Its $S_3$ automorphisms permute the three numerator factors. A transposition fixes the folded theory $A$, a three-cycle fixes the minimal model $C=W_3(5,6)$, and the full group fixes the bottom node $D=B^{S_3}$. The three transpositions give different copies of the same $A$ in $B$. Schematically, we have
\[
 \begin{array}{ccccc}
 &&B&&\\[-3pt]
 &\nearrow&&\nwarrow&\\[-3pt]
 C=W_3(5,6)&&&&A=\mathrm{Ising}\otimes\mathrm{TIM}\,,\\[-3pt]
 &\nwarrow&&\nearrow&\\[-3pt]
 &&D&&
 \end{array}
\]
where arrows point from subalgebra to extension. Theory $D$ is the bottom of this diagram; the search tests whether any rational restriction leads to an algebra outside it.\footnote{$C$ is not a subalgebra of $A$. One can get $C$ from $A$ by gauging, see Fig.~\ref{fig:web}}

\textbf{The upward list is exhaustive.} The only integer-weight irreducible $D$-modules are $\one,z,J$, with lowest weights $0,3,2$ and quantum dimensions $1,1,2$. Extending $D$ with these modules leads to all other nodes in the web; we will discuss this in more detail in later sections.

\textbf{The downward search gives bounds.} 
Theory $B$ is generated by $\Ttot$ and \(Y=\psi_{\rm I}G_{\rm TIM}\)~\cite{Gaiotto:2012np}.
The standard embedding of $A$ is generated by $\Ttot$ and  $7T_{\rm I}-5T_{\rm TIM}$.
The weight-three and weight-six primaries generate $C$ and $D$, respectively. 
All other possible values for the lightest weight through 21 are excluded in $B$, either by the character or by the OPE test. A separate calculation inside $D$ excludes further lightest weights through 24. Thus
\[
 \boxed{\begin{aligned}
 U\subseteq B\text{ of a new rational VOA}:&\quad h_*(U)\ge22,\\
 U\subsetneq D\text{ admissible and rational}:&\quad h_*(U)\ge25.
 \end{aligned}}
\]
If the four-node web is complete, $D$ is its unique bottom. 
Completeness, however, remains open, since as mentioned above, there could be other disconnected webs that we cannot access through this procedure. 
 
 \medskip

\paragraph{Comments on the generality of the \emph{moving up and down} approach}
Let us briefly discuss the relevance of the moving up and down approaches in the broad program of classifying RCFTs. Roughly speaking, all known examples of RCFTs can be obtained either by coset models, at special points in Narain CFTs moduli space or products and orbifolds thereof \cite{Yin:2017yyn}. Our approach gives a more systematic way of scanning the space of RCFTs at a given central charge, once one is found. We should \emph{move up}, i.e. try to extend the vacuum algebra by simple and non-simple current extensions, and \emph{move down} i.e. find rational subalgebras, be it by orbifolding or by the more general search procedure established above. As already mentioned above, this procedure is certainly not exhaustive, as it is restricted to a given connected web, of which there could be many.

A complementary approach to RCFT classification (which is also not exhaustive, due to an unbounded search space), can be broadly defined as the \emph{modular approach}. In the classic formulation of Mathur, Mukhi and Sen, one searches for characters by solving modular linear differential equations, organized by the number of independent characters and the Wronskian index~\cite{MathurMukhiSen1988,MathurMukhiSen1989}.
A complementary starting point is to enumerate the possible modular data $(S,T)$ of modular tensor categories, as in the work of Ng, Rowell, Wang and Wen and its extension to rank twelve~\cite{NgRowellWangWen2023,NgRowellWen2025}.
Once $(S,T)$ and the central charge $c$ are fixed, we can ask which sets of characters have a unique vacuum and nonnegative integer $q$-expansion coefficients~\cite{BantayGannon2006}.
Su has reported a specialized integer scan for this problem at high rank~\cite{SuBootstrap2026}.
However, even after fixing the characters, we still have to determine which chiral algebras realize them.

Used together, our \emph{moving up and down} and the \emph{modular} approaches give a very powerful algorithm to construct extensive sets of RCFTs and potentially challenge the lore on the limited set of origins these CFTs can have. It would be particularly interesting to consider the space of CFTs with central charge slightly above 1, recently discussed in \cite{Poland2026,Nakayama2026,KousvosPiazzaVichi2026,BenjaminLanghoffLiThaler2026}.

\section{The tetrahedral duality web}\label{sec:common}

The two constructions - moving up and moving down - use the same physical idea: keep enough chiral symmetry to calculate the boundary state, while allowing the Ising and TIM stress tensors to mix. We begin by locating the product theory inside the auxiliary coset and identifying the currents that survive its permutation gluings.

\subsection{Folding and the chiral extension}

Fold the TIM side of the junction and write
\begin{align}
 A=\mathcal M_{3,4}\boxtimes\overline{\mathcal M}_{4,5},
 \qquad c_A=\frac12+\frac7{10}=\frac65\,,
 \label{eq:A}
\end{align}
where $\mathcal M_{p,p+1}$ is a unitary Virasoro minimal model and the bar records the orientation reversal. After folding, the original TIM antiholomorphic fields become holomorphic fields of $A$. The conformal weights are unchanged, and the two central charges add.

An interface is now a boundary state $\ket R_A$. Conservation of energy at the junction becomes 
\begin{align}
 \left(L_n^{\Is}+\bar L_n^{\TIM}
       -\bar L_{-n}^{\Is}-L_{-n}^{\TIM}\right)\ket R_A=0\,,
 \qquad n\in\mathbb Z\,,
 \label{eq:totalgluing}
\end{align}
where the unbarred and barred $L_n$ retain their holomorphic and antiholomorphic labels before folding, with the superscripts identifying Ising or TIM. This equation glues the total stress tensor of the folded theory. Gluing each stress tensor separately would give two independent boundaries and total reflection. For transmission, we need a boundary that redistributes energy between the two factors while preserving the sum.

The coset used by Gaiotto provides such a boundary condition~\cite{Gaiotto:2012np}. It is
\begin{align}
 B=\frac{\widehat{\mathfrak{su}}(2)_1^{(1)}
          \times\widehat{\mathfrak{su}}(2)_1^{(2)}
          \times\widehat{\mathfrak{su}}(2)_1^{(3)}}
         {\widehat{\mathfrak{su}}(2)_3}\,,
 \label{eq:B}
\end{align}
where the denominator is the diagonal algebra of the three numerator factors. To recover Ising and TIM, combine factors one and two first:
\begin{align}
 \frac{\widehat{\mathfrak{su}}(2)_1^{(1)}
          \times\widehat{\mathfrak{su}}(2)_1^{(2)}}
         {\widehat{\mathfrak{su}}(2)_2^{(12)}}
 \boxtimes
 \overline{\frac{\widehat{\mathfrak{su}}(2)_2^{(12)}
          \times\widehat{\mathfrak{su}}(2)_1^{(3)}}
         {\widehat{\mathfrak{su}}(2)_3}}
 \ \subset B\,.
 \label{eq:AB}
\end{align}
The first coset has central charge $\frac12$ and the second has central charge $\frac7{10}$. Their product has the same stress tensor as $B$. The extra sector in the $B$ vacuum combines the Ising primary of weight $\frac{1}{2}$ with the TIM primary of weight $\frac{3}{2}$ into an additional current of spin-two 
\begin{align}
 X\propto\phi^{\Is}_{1,3}\otimes
 \overline{\phi^{\TIM}_{3,1}}\,,
 \qquad h_X=\frac{1}{2}+\frac{3}{2}=2\,,
 \label{eq:Bphantom}
\end{align}
where the subscripts are the minimal model Kac labels and the bar denotes the TIM antiholomorphic field before folding. The current $X$ is holomorphic after folding and extends the product chiral algebra to $B$.

All three numerator factors in the coset Eq.~\eqref{eq:B} are at level one, so they can be permuted while leaving theory $B$ invariant. We denote the exchange of numerator factors $i$ and $j$ by $\sigma_{ij}$, and the 3-cycle permutation by $\sigma_{ijk}$. The exchange $\sigma_{12}$ fixes the two physical stress tensors $T_{\Is}$ and $T_{\TIM}$, while sending the extra current to $\sigma_{12}(X)=-X$. Gauging the $\mathbb Z_2$ symmetry generated by $\sigma_{12}$ therefore recovers the physical Ising and TIM product theory $A$, whose chiral algebra consists of the $\sigma_{12}$ invariant fields in $B$. The exchange $\sigma_{23}$ is used in the RG boundary of \cite{Gaiotto:2012np}. The other nodes of the web are the invariant subalgebras
\begin{align}
 A=B^{\langle\sigma_{12}\rangle},\qquad
 C=B^{\langle\sigma_{123}\rangle}=W_3(5,6),\qquad
 D=B^{S_3}=C^{\mathsf{cc}}\,,
 \label{eq:web}
\end{align}
where $\mathsf{cc}$ is charge conjugation of the $W_3$ algebra. These descriptions can also
be obtained from the common lattice realization of the cosets~\cite{Dong_2004,Dong_2017}.
The four theories have central charge $\frac65$, and their numbers of irreducible representations are $18$, $8$, $20$ and $24$ in the order $A,B,C,D$. Gauging and extending between them gives Figure~\ref{fig:web}.

\subsection{The currents preserved at the interface}

To determine which chiral symmetries survive the pullback to $A$, consider a boundary of $B$ whose gluing is twisted by $\sigma_{23}$~\cite{Gaiotto:2012np}. This transposition exchanges numerator factors $2$ and $3$, replacing the pair $(12)$ used to construct Ising by $(13)$. The resulting embedding of the Ising and TIM product in $B$ is denoted by $\sigma_{23}(A)$. At the boundary, a holomorphic current of $B$ is glued to the antiholomorphic counterpart of its $\sigma_{23}$ image. This gluing can be carried back to the physical theory when both the current and its image belong to the chiral algebra $A$. Since $\sigma_{23}^2=1$, these currents form $A\cap\sigma_{23}(A)$. Using $A=B^{\langle\sigma_{12}\rangle}$ and $\sigma_{23}\sigma_{12}\sigma_{23}^{-1}=\sigma_{13}$, we obtain
\begin{align}
 A\cap\sigma_{23}(A)
 =B^{\langle\sigma_{12}\rangle}\cap B^{\langle\sigma_{13}\rangle}
 =B^{S_3}=D\,.
 \label{eq:commonalgebra}
\end{align}
The two transpositions generate $S_3$, which
explains why its invariant algebra appears.
Every permutation fixes $D$, so the boundaries
constructed in $B$ all preserve ordinary
gluing of $D$ after the pullback.

For the spin-two currents this statement has a direct
 interpretation. Denote the Sugawara
stress tensors of numerator factor $i$, the
diagonal pair $ij$ and the full denominator
by $T_i$, $T_{ij}$ and $T_{123}$. The physical
stress tensors are
\begin{align}
 T_{\Is}&=T_1+T_2-T_{12},&
 T_{\TIM}&=T_{12}+T_3-T_{123}\,.
 \label{eq:embeddedstress}
\end{align}
The exchange $\sigma_{23}$ replaces $T_{12}$ by $T_{13}$,
mixing these energy densities with the extra
current of spin two. Their sum stays fixed:
\begin{align}
 T_{\mathrm{tot}}=T_{\Is}+T_{\TIM}
 =T_1+T_2+T_3-T_{123}\,.
 \label{eq:totalstress}
\end{align}
The three dimensional weight two space of
the $B$ vacuum consists of this invariant
direction and an $S_3$ doublet. The doublet
contains the combinations that are mixed at
the interface.

Higher currents can also survive the gluing.
They are particularly easy to see in
$C=W_3(5,6)$, which is generated by the stress
tensor and a primary $W$ of spin three.
Charge conjugation acts as
\begin{align}
 \mathsf{cc}(T_{\mathrm{tot}})=T_{\mathrm{tot}},\qquad
 \mathsf{cc}(W)=-W\,.
 \label{eq:cc}
\end{align}
Even normal-ordered products of $W$ survive in $D$. Removing Virasoro descendants from the normal-ordered product $:WW:$ gives a primary $U_6$ of weight six~\cite{Al-Ali:2016ndc}. With $q=e^{2\pi i\tau}$, $\operatorname{Im}\tau>0$, the vacuum characters begin as
\begin{align}
 q^{\frac{1}{20}}\chi_{\mathrm{Vir}_{\frac{6}{5}},0}(q)
 &=1+q^2+q^3+2q^4+2q^5+4q^6+O(q^7)\,,\notag\\
 q^{\frac{1}{20}}\chi_0(q)
 &=1+q^2+q^3+2q^4+2q^5+5q^6+O(q^7)\,,
 \label{eq:vacuumcomparison}
\end{align}
where the common factor $q^{-\frac{1}{20}}$ is the
vacuum energy contribution at $c=\frac{6}{5}$.
The additional state at weight six is $U_6$.
The boundary therefore conserves this current
together with the total stress tensor.
Its operator spectrum can be grouped into
representations of the rational algebra $D$.

The OPEs of $U_6$ determine the
remaining currents and their possible gluings.
Normalize $W$ by
$\langle W(z)W(0)\rangle=(\frac{2}{5})z^{-6}$ and
choose the coefficient of $W_{-3}^2|0\rangle$
in $U_6$ to be one.
The even part of the universal $W_3$ algebra
has generators of weights $2,6,8,10,12$
under normal products and derivatives~\cite{Al-Ali:2016ndc}.
In the minimal model, the weight twelve
primary is already generated by $:U_6U_6:$,
while the weight eight and ten primaries
appear in the singular terms of
$U_6(z)U_6(0)$. Thus the stress tensor
and $U_6$ generate $D$ through their
OPE. In our normalization,
\begin{align}
 \langle U_6(z)U_6(0)\rangle
 &=\frac{789888}{233975}\,z^{-12},\qquad
 U_6(z)U_6(0)\supset
 -\frac{1097536}{46795}\frac{U_6(0)}{z^6}\,.
 \label{eq:U6coupling}
\end{align}
An automorphism could only multiply the
unique weight six primary by a number.
Preserving its two point function restricts
that number to a sign, and the nonzero
self coupling fixes the sign to be positive.
Since the other currents are generated
from OPEs of $U_6$ and the stress tensor, all of
them are fixed as well. Hence
$\operatorname{Aut}_{\mathrm{conf}}(D)=\{1\}$.

A smaller rational algebra would have to
omit $U_6$ and all currents whose operator
products regenerate it. 
For a proper strongly rational subalgebra
with the same stress tensor, we are able to show that the first
additional primary must have weight at leasttwenty-five, using the algorithm discussed in Section \ref{downmove}.
We therefore conjecture that no smaller rational subalgebra exists. 

\subsection{Extending the common algebra back to the product}

To turn a $D$ boundary into an Ising and TIM interface, we restore the fields that extend $D$ to $A$. Write $D_h$ for an irreducible $D$ module of lowest weight $h$, and $\chi_h(\tau)=\tr_{D_h}q^{L_0-\frac{1}{20}}$ for its character. The 24 primaries have distinct conformal weights. Among them, three modules describe the extensions in the web: the vacuum $\one=D_0$, $z=D_3$ and $J=D_2$. Their fusion rules are
\begin{align}
 z\otimes z=\one,\qquad z\otimes J=J,\qquad
 J\otimes J=\one\oplus z\oplus J\,,
 \label{eq:electric}
\end{align}
These are the rules for the trivial, sign
and two-dimensional representations of
$S_3$. The corresponding conformal embeddings are
\begin{align}
 D\xhookrightarrow{\ \one\oplus z\ }C,\qquad
 D\xhookrightarrow{\ \one\oplus J \ }A,\qquad
 D\xhookrightarrow{\  \one\oplus z\oplus2J \ }B\,.
 \label{eq:vacbranch}
\end{align}
In particular, recovering the physical tensor
product theory $A$ requires adjoining $J$. The
fields in $\one\oplus J$, with their
inherited multiplication, form the
Frobenius algebra~\cite{Fuchs:2002cm,Bhardwaj:2017xup}
\begin{align}
 \fA=\one\oplus J
 \simeq\operatorname{Fun}\bigl(S_3/\langle\sigma_{12}\rangle\bigr),
 \qquad \qdim(\fA)=3\,,
 \label{eq:FA}
\end{align}
where $\operatorname{Fun}(S_3/\langle\sigma_{12}\rangle)$ is the algebra of functions on the three cosets of the subgroup generated by $\sigma_{12}$. This algebra specifies which fields become part of the extended vacuum and how they combine. The associated topological interface implements the extension at the level of boundary states~\cite{Diatlyk:2023fwf}. Gauging and extension relate both the bulk sectors and their boundary conditions~\cite{Fukusumi:2024ejk}.

We also need to know how the physical bulk
states decompose under the preserved symmetry.
A product representation $A_a$, labelled
by $a=(x,y)$, has chiral weights
\begin{align}
 x&\in\left\{0,\frac1{16},\frac12\right\},&
 y&\in\left\{0,\frac1{10},\frac35,\frac32,
                 \frac3{80},\frac7{16}\right\}\,,
 \label{eq:Alabels}
\end{align}
in Ising and TIM, respectively. Define
the branching matrix $b$ by
\begin{align}
 \Res_D A_a=\bigoplus_i b_{ai}D_i,\qquad
 \chi^A_a=\sum_i b_{ai}\chi_i\,,
 \label{eq:branch}
\end{align}
where $i$ runs over the twenty four $D$
labels listed in Appendix~\ref{app:modular}.
Thus $b$ has eighteen rows and twenty
four columns, and $b_{ai}$ counts the
copies of $D_i$ inside $A_a$.
For example, the product vacuum gives
$D_0\oplus D_2$, while the sector
$(\frac{1}{2},\frac{3}{2})$ gives $D_2\oplus D_3$.
Six product representations split into
two $D$ modules. Each of the remaining
twelve becomes the single module
$D_{x+y}$, as listed in
Table~\ref{tab:branching} in Appendix~\ref{app:modular}.

A boundary state must be assembled from
the bulk states of the physical product theory.
Expressing its partition function in
$D$ characters gives
\begin{align}
 Z_A(\tau,\bar\tau)
 &=\sum_a\left|\sum_i b_{ai}\chi_i(\tau)\right|^2
   =\sum_{i,j}\mathsf Z_{ij}
         \chi_i(\tau)\overline{\chi_j(\tau)},&
 \mathsf Z&=b^{\mathsf T}b\,.
 \label{eq:ZA}
\end{align}
The matrix $\mathsf Z$ takes into account both the available representations and their
multiplicities. For ordinary $D$ gluing, the number of Ishibashi states is therefore
\begin{align}
 \sum_i\mathsf Z_{ii}=24\,.
 \label{eq:Ishcount}
\end{align}
These twenty four Ishibashi states are the ingredients for the elementary boundaries constructed next.

\section{Non-factorizable interfaces from \texorpdfstring{$D$}{D} and \texorpdfstring{$B$}{B}}
\label{sec:construction}

%\JC{sec 4 is done}
%\ALA{I am happy with this section (only touched the grammar)}
%\JR{all good}

We now construct non-factorizable boundary states of the theory $A=\text{Ising$\,\times\,$TIM}$ from both the moving down and up approaches. In the former case, we start with $24$ ordinary Cardy states in $D$, and pull them back to A via the topological interface $\mathcal I_{D\to A}$ defined in the tetrality web of Figure~\ref{fig:web}. Alternatively and as a consistency check, we also construct twisted Cardy states in $B$ and map them back to $A$ by the topological interface $\mathcal I_{B\to A}$. We then find the same $6$ non-factorizable branes identified by the two different routes.

\subsection{Moving down: from ordinary Cardy states in \texorpdfstring{$D$}{D}}
\label{sec:Dprojector}

We prepare a boundary state in the diagonal theory of $D$ and bring the topological interface $\cI_{D\to A}$ onto it to obtain a boundary state in $A$. The currents in the vacuum module of $D$ can pass through this interface to $A$, so their gluing is preserved when the bulk becomes the physical Ising and TIM product. These operators can couple differently to the boundary, and thus allows the two stress tensors in Ising and TIM to mix.

Let $i,j$ label the $24$ irreducible modules of $D$ in Table~\ref{tab:Dmodules} in Appendix~\ref{app:modular}, with $0$ denoting the vacuum. Ordinary gluing gives an Ishibashi state $\ish{j}_D$ corresponding to each module~\cite{Ishibashi:1988kg}. We normalize these states by
\begin{align}
 {}_D\langle\!\langle i|\tilde q^{H_c}|j\rangle\!\rangle_D
 =\delta_{ij}\chi_i(\tilde q)\,,\qquad
 H_c=\frac{L_0+\bar L_0-\frac{c}{12}}{2}\,,\quad\text{with}\quad c=\frac65\,,
 \label{eq:DIshnorm}
\end{align}
where $\tilde q$ is the closed channel nome. The ordinary Cardy
states then spell as~\cite{Cardy:1989ir}
\begin{align}
 \ket{i}_D
 =\sum_j\frac{S^D_{ij}}{\sqrt{S^D_{0j}}}\ish{j}_D\,,
 \label{eq:DCardy}
\end{align}
where $S^D$ is the modular matrix in Eq.~\eqref{eq:SD}. To transport the state, we apply the branching of a physical product module $A_\alpha$ into $D_j$ modules, as shown in Table~\ref{tab:branching}, where a label $\alpha=(x,y)$ records the Ising and TIM weights in Eq.~\eqref{eq:Alabels}. Each such branching will assign a chiral embedding and its projector,
\begin{align}
 &e_{\alpha j}:D_j\longrightarrow A_\alpha\,,
 \qquad P_{\alpha j}=e_{\alpha j}e_{\alpha j}^{\dagger}\,,
 \notag\\[5pt]
 &e_{\alpha j}^{\dagger}\,e_{\alpha k}
 =\delta_{jk}\one_{D_j}\,,
 \qquad
 \sum_{j:\,b_{\alpha j}=1}P_{\alpha j}=\one_{A_\alpha}\,,
 \label{eq:Dembeddings}
\end{align}
where the chiral embedding $e_{\alpha j}$ is determined by the branching Eq.~\eqref{eq:branch}, and $e_{\alpha j}^{\dagger}$ is the dual of $e_{\alpha j}$ from the module $A_\alpha$ restricted onto $D_j$. These maps act on the highest-weight states and all their descendants. For instance, for the highest weight $\ket2_D^{\mathrm{hw}}$ in module $D_2$, $e_{(0,0),2}$ embeds it to the relative stress-tensor of Ising and TIM,
\begin{align}
 e_{(0,0),2}\ket2_D^{\mathrm{hw}}
 =\frac{7L^{\Is}_{-2}-5L^{\TIM}_{-2}}{\sqrt{21}}
   \bigl(\ket0_{\Is}\otimes\ket0_{\TIM}\bigr)\,.
 \label{eq:D2embedding}
\end{align}
Using the embedding and projector in Eq.~\eqref{eq:Dembeddings}, we define
\begin{align}
 \ish{\alpha;j}_A
\equiv(e_{\alpha j}\otimes\overline{e_{\alpha j}})\ish{j}_D
   =(P_{\alpha j}\otimes\overline{\one})\ish{\alpha}_A\,,
\end{align}
where $\ish{\alpha}_A=\ish{x}_{\Is}\otimes\ish{y}_{\TIM}$ is the ordinary product Ishibashi state, and the second equality is obtained by noticing the Ishibashi states are diagonally gluing the chiral and anti-chiral pieces. In terms of $\ish{\alpha;j}_A$, an Ishibashi state in $A$ can be written as
\begin{align}
\ish{\alpha}_A
 =\sum_{j:\,b_{\alpha j}\neq 0}\ish{\alpha;j}_A\,,
 \label{eq:DphysicalIsh}
\end{align}

The topological interface carries each diagonal $D$ sector into
all its copies in the $A$ theory, with additional normalizations as explained in~\cite{Gaiotto:2012np,Chen:2026byx}
\begin{align}
 \cI_{D\to A}\ish{j}_D
 =\sum_{\alpha:\,b_{\alpha j}\neq0}
   \rho_{\alpha j}\ish{\alpha;j}_A\,,
   \qquad
 \rho_{\alpha j}
 =\sqrt{\frac{S^A_{0\alpha}}{S^D_{0j}}}\,,
 \label{eq:Dinterfaceaction}
\end{align}
where $S^A=S^{\Is}\otimes S^{\TIM}$. Thus its action on a complete
Cardy state of the theory $D$ gives
\begin{align}
 \ket{\mathcal B_i}_A
\equiv\cI_{D\to A}\ket{i}_D
 &=\sum_{\alpha,j:\,b_{\alpha j}\neq 0}
   \sqrt{S^A_{0\alpha}}\,
   \frac{S^D_{ij}}{S^D_{0j}}\ish{\alpha;j}_A
 \notag\\[5pt]
 &=\sum_\alpha\sqrt{S^A_{0\alpha}}
   \left(
    \left(\sum_{j:\,b_{\alpha j}\neq0}
      \frac{S^D_{ij}}{S^D_{0j}}P_{\alpha j}\right)
      \otimes\overline{\one}
   \right)\ish{\alpha}_A\,.
 \label{eq:Dtransport}
\end{align}
The brane $\ket{\mathcal B_i}_A$ is usually not a simple object, but rather a direct sum of (non)-factorized boundary states in $A$. To isolate them from each other, we first identify the fully reflective components. Denote the ordinary product Cardy boundary with label $a=(x,y)$ by $\mathcal F_a\equiv\mathcal F_{x,y}$. It has the expansion
\begin{align}
 \ket{\mathcal F_{x,y}}_A
 &=\sum_{u,v}
 \frac{S^{\Is}_{xu}S^{\TIM}_{yv}}
 {\sqrt{S^{\Is}_{0u}S^{\TIM}_{0v}}}
 \ish{u}_{\Is}\otimes\ish{v}_{\TIM}\,,
 \label{eq:localCardy}
\end{align}
These $18$ boundaries glue the two stress tensors separately and reflect all incident energy. In particular, for the vacuum label $i=0$, one simply finds that $\ket{\mathcal B_0}_A=\ket{\mathcal F_{0,0}}_A$. The multiplicity of a brane $\mathcal F_{a}$ in $\mathcal B_i$ is measured by their mixed annulus amplitude, denoted by $n_0(\mathcal B_i,\,\mathcal F_a)$. Fixing an elementary boundary, this number counts how many copies of that boundary occur in the other end of the strip. From Eq.~\eqref{eq:Dtransport} and \eqref{eq:localCardy}, we precisely have
\begin{align}
 n_0(\mathcal B_i,\,\mathcal F_a)
 &=\sum_{\alpha,j}S^A_{a\alpha}\,b_{\alpha j}\,S^D_{ij}
   =(S^A\cdot b\cdot S^D)_{ai}=b_{ai}\,,
 \label{eq:Dlocalmultiplicity}
\end{align}
where we have used the interface property $b\cdot S^D=S^A\cdot b$ and $(S^A)^2=\one$. We can therefore remove the reflecting components and define
\begin{align}
 \ket{\mathcal R_i}_A
 \equiv\ket{\mathcal B_i}_A-\sum_a b_{ai}\ket{\mathcal F_a}_A\,.
 \label{eq:Dremainder}
\end{align}
Further notice that 
\begin{align}
    n_0(\mathcal B_i,\,\mathcal B_j)=\left\langle D_i\right|\mathcal I_{A\to D}\circ\mathcal I_{D\to A}\left|D_j\right\rangle=\delta_{ij}+(N_J)_{ij}\,,
\end{align}
where we have used the facts,
\begin{align}
    \mathcal I_{A\to D}\circ\mathcal I_{D\to A}=\mathcal L_\one\oplus \mathcal L_J\,,\qquad
    (N_J)_{ij}=\left\langle D_i\right|\mathcal L_J\left|D_j\right\rangle\,.
\end{align}
Here $\mathcal L_{\one}\oplus\mathcal L_J$ are the Verlinde lines corresponding to the modules $D_\one$ and $D_J$ in $D$, and $(N_J)_{ij}$ counts the multiplicity of $D_j$ in $J\otimes D_i$. The remaining branes $\mathcal R_i$ have their mixed annulus amplitude given by
\begin{align}
    n_0(\mathcal R_i,\,\mathcal R_j)=\delta_{ij}+(N_{J})_{ij}-(b^{\mathsf T}b)_{ij}=\text{diag}\left(0_6,\,E_3,\,E_3,\,0_4,\,E_2,\,E_2,\,E_2,\,E_2\right)\,,
    \label{eq:Dresidualvacua}
\end{align}
where $0_n$ denotes a $n\times n$ null matrix and $E_n$ a $n\times n$ matrix with every entry equal to $1$. The $6$ $E_n$ blocks in Eq.~\eqref{eq:Dresidualvacua} indicate that there are 6 non-factorizable conformal interfaces pulled back from the ordinary Cardy states in the theory $D$.

To demonstrate the procedure explicitly, for example, consider a pair of Cardy state labelled by $j=\frac{1}{16}$ and $\frac{25}{16}$ in $D$. After pullback and removing the reflective components, we end up with two identical branes 
\begin{align}
    \ket{\mathcal T_1}_A\equiv\ket{\mathcal R_{\frac{1}{16}}}_A=\ket{\mathcal R_{\frac{25}{16}}}_A\,,
\end{align}
furnishing an $E_2$ block. More concretely, one can find that 
\begin{align}
    &\ket{\mathcal B_{\frac{1}{16}}}_A=\ket{\mathcal F_{\frac{1}{16},0}}_A+\ket{\mathcal T_1}_A\,,\qquad {\rm and}\qquad
    \ket{\mathcal B_{\frac{25}{16}}}_A=\ket{\mathcal F_{\frac{1}{16},\frac32}}_A+\ket{\mathcal T_1}_A\,,\notag\\[7pt]
    \text{with}\quad
    & \ket{\mathcal T_1}_A=\sum_\alpha\sqrt{2S^A_{0\alpha}}
   \left(
    \left(\,\sum_{j:\,b_{\alpha j}\neq 0}
       \kappa_{\alpha j}P_{\alpha j}\right)
      \otimes\overline{\one}
   \right)\ish{\alpha}_A\,,
   \label{eq:T0Dprojectors}
\end{align}
where $\kappa_{\alpha j}$ are computed in Table~\ref{tab:T0Dprojectors}. The different coefficients on the total and relative stress tensors in Eq.~\eqref{eq:D2embedding} allow the Ising and TIM stress tensors to mix at this boundary.

\begin{table}[tbp]
 \centering
 \begin{tabular}{c c c}
 \toprule
 $A$ module $\alpha=(x,y)$ & $D$ module $j$ &
 $\kappa_{\alpha j}$\\
 \midrule
 $(0,0)$ & $(0,2)$ & $(2,-1)$\\
 $\left(0,\frac35\right)$ & $\left(\frac{13}{5},\frac35\right)$ & $(2,-1)$\\
 $\left(\frac{1}{16},\frac{7}{16}\right)$ & $\left(\frac{1}{2},\frac32\right)$ & $(-1,1)$\\
 $\left(\frac{1}{16},\frac{3}{80}\right)$ & $\left(\frac{11}{10},\frac{1}{10}\right)$ & $(-1,1)$\\
 $\left(\frac{1}{2},\frac{3}{2}\right)$ & $(2,3)$ & $(1,-2)$\\
 $\left(\frac{1}{2},\frac{1}{10}\right)$ & $\left(\frac{3}{5},\frac{8}{5}\right)$ & $(1,-2)$\\
 \bottomrule
 \end{tabular}
 \caption{The projector coefficients of $\mathcal T_1$. All other unlisted coefficients vanish.}
 \label{tab:T0Dprojectors}
\end{table}

In the next subsections~\ref{sec:Bproject} and \ref{sec:identifyBD}, we will identify $\mathcal T_1$ with the RG brane between Ising and TIM constructed in~\cite{Gaiotto:2012np}. For now, we label the $6$ $E_n$ blocks for $6$ distinct conformal interfaces $\mathcal T_1$, $\mathcal T_\tau$, $\mathcal T_\eta$, $\mathcal T_{\eta\tau}$, $\mathcal T_\sigma$, and $\mathcal T_{\sigma\tau}$. Their subscripts will become clear in Section~\ref{sec:spectra}. Pushing them forward to the theory $D$ via $\mathcal I_{A\to D}$, we obtain the following sums of ordinary Cardy states in $D$:
\begin{align}
\mathcal I_{A\to D}|\mathcal T_1\rangle_A
&=\left|\frac{1}{16}\right\rangle_D
 +\left|\frac{25}{16}\right\rangle_D\,,
 &\quad\quad
\mathcal I_{A\to D}|\mathcal T_\tau\rangle_A
&=\left|\frac{13}{80}\right\rangle_D
 +\left|\frac{53}{80}\right\rangle_D\,,
\label{eq:sixrestrictions}
\\[5pt]
\mathcal I_{A\to D}|\mathcal T_\eta\rangle_A
&=\left|\frac{7}{16}\right\rangle_D
 +\left|\frac{15}{16}\right\rangle_D\,,
&
\mathcal I_{A\to D}|\mathcal T_{\eta\tau}\rangle_A
&=\left|\frac{3}{80}\right\rangle_D
 +\left|\frac{43}{80}\right\rangle_D\,,
\notag\\[5pt]
\mathcal I_{A\to D}|\mathcal T_\sigma\rangle_A
&=\left|\frac{1}{9}\right\rangle_D
 +\left|\frac{7}{9}\right\rangle_D
 +\left|\frac{13}{9}\right\rangle_D\,,
&
\mathcal I_{A\to D}|\mathcal T_{\sigma\tau}\rangle_A
&=\left|\frac{2}{45}\right\rangle_D
 +\left|\frac{17}{45}\right\rangle_D
 +\left|\frac{32}{45}\right\rangle_D\,.\notag
\end{align}

\subsection{Moving up: from twisted Cardy states in \texorpdfstring{$B$}{B}}
\label{sec:Bproject}

We now use the moving up approach to construct conformal interfaces between Ising and TIM. This construction was introduced by Gaiotto~\cite{Gaiotto:2012np} and recently revisited and developed in~\cite{Chen:2026byx}. The idea is to pass from the physical theory $A$ to the auxiliary theory $B$, which has the largest chiral algebra in the tetrality web of Figure~\ref{fig:web}. We prepare twisted Cardy states in $B$, and pull them back to $A$ through the topological interface $\mathcal I_{B\to A}$. The mechanism that allows energy transmission on the interface is the condensation of the phantom current $X$ of spin two introduced in Eq.~\eqref{eq:Bphantom}. In the folded theory $A$, the diagonal Verlinde $\mathbb Z_2$-line\,,
\begin{align}
    j_A\equiv \mathcal L_{\frac12}^\Is\boxtimes\overline{\mathcal L_{\frac32}^\TIM}\,,
    \label{eq:jA}
\end{align}
has $X$ as its chiral endpoint field. Gauging this line extends the chiral algebra $A$ to $B$, as in Eq.~\eqref{eq:A_to_B}, and brings $X$ into the auxiliary theory vacuum. Writing $B_0$ for that vacuum module, its restriction onto $A$ and the full subspace of weight-two are
\begin{align}
 \left.B_0\right|_A
 =A_{(0,0)}\oplus A_{(\frac{1}{2},\frac{3}{2})}\,,
 \qquad
 (B_0)_{h=2}
 =\operatorname{span}\left\{\,T_{\Is},\,T_{\TIM},\,X\,\right\}\,.
 \label{eq:BvacuumMixing}
\end{align}
Thus the two stress tensors can mix with $X$ at a boundary while their sum remains conserved~\cite{Furuta:2025ahl}. Following~\cite{Chen:2026byx}, we implement this mixing by an automorphism $\Omega$ of the full chiral algebra $B$, which thus can be used to define an $\Omega$-twisted Cardy state in $B$.

For the coset at hand (Eq.~\eqref{eq:B}), the automorphism group is $S_3$, acting by permutations of its three numerator factors. The identity and transposition $\sigma_{12}$ fix $T_{\Is}$ and $T_{\TIM}$ separately, with $\sigma_{12}(X)=-X$. The other four permutations
\begin{align}
 g\in\mathcal S_\sigma\equiv\{\sigma_{23},\,\sigma_{13},\,\sigma_{123},\,\sigma_{132}\}
 \label{eq:transmittingPermutations}
\end{align}
instead mix the two stress tensors and $X$, providing the non-trivial gluings needed for establishing non-factorizable interfaces in the physical theory $A$. To define a twisted Cardy state, one further needs to specify a Cardy label $\lambda$ for a given twist $g$. Notice that in $B$, there are $8$ irreducible modules labeled by $\mu=[a,b,c;d]$, where $a,b,c\in\{1,2\}$ label the level one numerator $\widehat{\mathfrak {su}}(2)_1$ and $d\in\{1,2,3,4\}$ labels the diagonal $\widehat{\mathfrak {su}}(2)_3$. The $\mu=[a,b,c;d]$ are further subject to the following selection rules and equivalence relation
\begin{align}
 a+b+c+d\in2\mathbb Z\,,
 \qquad
 [a,b,c;d]\simeq[3-a,3-b,3-c;5-d]\,,
 \label{eq:Blabels}
\end{align}
and thus label the 8 promised modules. The modular matrix of $B$ can be assembled from the modular matrix $S^{(k)}$ for the affine $\widehat{\mathfrak {su}}(2)_k$, i.e.
\begin{align}
 S^{(k)}_{uv}
 =\sqrt{\frac{2}{k+2}}\sin\frac{\pi uv}{k+2}\,,\qquad
 S^B_{\mu\nu}
 =2S^{(1)}_{aa'}\,S^{(1)}_{bb'}\,S^{(1)}_{cc'}\,S^{(3)}_{dd'}\,,
 \label{eq:SB}
\end{align}
for the labels $\mu=[a,b,c;d]$ and $\nu=[a',b',c';d']$, where the additional prefactor $2$ is due to the field identification of the coset construction~\cite{Ishikawa:2001zu}. Obviously, in $B$, one has $8$ ordinary Ishibashi states $\ish{\mu}_B$.

On the other hand, to define a $g$-twisted Ishibashi state, one has to first pick the $g$-invariant labels collected in the set
\begin{align}
 \mathcal P_g
 =\left\{\mu=[a,b,c;d]\ \middle|\ g\cdot\mu\simeq\mu\right\}\,,
    \label{eq:Pset}
\end{align}
where $g\cdot\mu$ permutes the three numerator factors and leaves the denominator label $d$ unchanged, with the equivalence understood as the field identification in Eq.~\eqref{eq:Blabels}. A $g$-twisted Ishibashi state is also supported on each $\mu\in\mathcal P_g$. The $g$-induced unitary map $\Omega_{g,\mu}$ implements $g$ on the chiral module $B_\mu$, and defines the corresponding twisted Ishibashi state,
\begin{align}
 \ish{\mu;g}_B
 =(\Omega_{g,\mu}\otimes\overline{\one})\ish{\mu}_B\,\quad \Longrightarrow\quad
 \left(g(W)_n-(-1)^h\bar W_{-n}\right)\ish{\mu;g}_B=0\,,
 \label{eq:twistedIsh}
\end{align}
satisfying the $g$-glued boundary condition, where $W$ are the currents of spin-$h$ in the chiral algebra $B$. Furthermore, a legitimate $g$-twisted  boundary state satisfying the Cardy condition has to take the form~\cite{Recknagel:2002qq,Ishikawa:2001zu}
\begin{align}
 \ket{b^g_\lambda}_B
 =\sum_{\mu\in\mathcal P_g}
 \frac{\Psi^g_{\lambda\mu}}{\sqrt{S^B_{0\mu}}}
 \ish{\mu;g}_B\,,
 \label{eq:seed}
\end{align}
where $0=[1,1,1;1]$ is the vacuum label, and $\lambda$ labels an irreducible module of $B$ with the twist $g$. We normalize these $g$-twisted Ishibashi states $\ish{\mu;g}_B$ so that their self-overlap give the ordinary character $\chi^B_\mu$. The matrix $\Psi^g$ relates the characters of these twisted modules to the twining characters of the ordinary modules in $\mathcal P_g$. More explicitly, with $q=e^{2\pi i\tau}$ and $c_B=\frac{6}{5}$, the twining character is
\begin{align}
 \chi^g_\mu(\tau)
 =\operatorname{Tr}_{B_\mu}
 \left(\Omega_{g,\mu}\,q^{L_0-\frac{c_B}{24}}\right)\,,
 \qquad \mu\in\mathcal P_g\,.
 \label{eq:BtwiningCharacter}
\end{align}
A modular transformation exchanges this symmetry insertion with
a twist around the spatial circle. Denoting the characters in
that twisted sector by $\chi^{\mathrm{tw},g}_\lambda$, we define
\begin{align}
 \chi^{\mathrm{tw},g}_\lambda(\tau)
 =\sum_{\mu\in\mathcal P_g}
 \Psi^g_{\lambda\mu}\,
 \chi^g_\mu(-\frac{1}{\tau})\,.
 \label{eq:BtwistedModular}
\end{align}
Consequently, the annulus amplitude between the ordinary vacuum Cardy state and $\ket{b^g_\lambda}_B$ leads to an open-channel  character $\chi^{\mathrm{tw},g}_\lambda$. For two boundaries with the same twist, the multiplicity in front of an ordinary module $B_\nu$
\begin{align}
 n^g_{\nu;\lambda\lambda'}
 =\sum_{\mu\in\mathcal P_g}
 \bigl(\Psi^g_{\lambda\mu}\bigr)^*
 \Psi^g_{\lambda'\mu}
 \frac{S^B_{\mu\nu}}{S^B_{0\mu}}\,,
 \label{eq:BtwistedCardyCondition}
\end{align}
needs to be a non-negative integer and in particular $n^g_{0;\lambda\lambda'}=\delta_{\lambda\lambda'}$.

Now we explicitly spell out various $g$-twisted Cardy states in $B$. First consider $g=\sigma_{23}$. We fix the representative and choose the $\sigma_{23}$-invariant set
\begin{align}
 \mathcal P_{\sigma_{23}}
 =\left\{[1,b,b;d]\ \middle|\ b=1,2,\ d=1,3\right\}\,.
 \label{eq:Ptransposition}
\end{align}
Therefore the same representatives,
\begin{align}
 \lambda\in
 \left\{(1,1;1),\ (1,1;3),\ (2,2;1),\ (2,2;3)\right\}\,,
 \label{eq:BtranspositionRows}
\end{align}
label the four $g$-twisted Cardy states. The modular transformation of their characters gives the $\Psi^{\sigma_{23}}$ matrix,
\begin{align}
 \Psi^{\sigma_{23}}_{(b,b;d),[1,b',b';d']}
 =2S^{(1)}_{b1}S^{(1)}_{bb'}S^{(3)}_{dd'}\,,
 \label{eq:Psit}
\end{align}
Similarly, for $g=\sigma_{123}$, the invariant labels instead reduce to
\begin{align}
 \mathcal P_{\sigma_{123}}
 =\left\{[1,1,1;d]\ \middle|\ d=1,3\right\}\,.
 \label{eq:Pcycle}
\end{align}
We again retain these representatives and label the $\sigma_{123}$-twisted Cardy states by
\begin{align}
  \lambda\in\left\{(1;1),\ (1;3)\right\}\,.
 \label{eq:BcycleLabels}
\end{align}
The corresponding modular matrix is
\begin{align}
 \Psi^{\sigma_{123}}_{(1;d),[1,1,1;d']}
 =2S^{(1)}_{11}S^{(3)}_{dd'}
 =\sqrt{2}\,S^{(3)}_{dd'}\,,\qquad d,d'=1,3\,.
 \label{eq:Psir}
\end{align}
It has two independent rows, represented by
\begin{align}
 \lambda\in\left\{(1;1),\ (1;3)\right\}\,,
 \label{eq:BcycleRows}
\end{align}
and therefore gives two elementary Cardy states with gluing
$\sigma_{123}$. Substituting Eqs.~\eqref{eq:Psit} and \eqref{eq:Psir} into Eq.~\eqref{eq:seed}, we obtain all four $\sigma_{23}$ states and both $\sigma_{123}$ states explicitly:
\begin{align}
 &\ket{b^{\sigma_{23}}_{(b,b;d)}}_B
 =2^{\frac{1}{4}}\sum_{d'=1,3}
 \frac{S^{(3)}_{dd'}}{\sqrt{S^{(3)}_{1d'}}}
 \left(\ish{111;d';\sigma_{23}}_B+
       (-1)^{b-1}\ish{122;d';\sigma_{23}}_B\right)\,,\notag\\[5pt]
 &\ket{b^{\sigma_{123}}_{(1;d)}}_B
 =2^{\frac{3}{4}}\sum_{d'=1,3}
 \frac{S^{(3)}_{dd'}}{\sqrt{S^{(3)}_{1d'}}}
 \ish{111;d';\sigma_{123}}_B\,,
 \label{eq:referenceSeeds}
\end{align}
Finally, notice that the conjugation action by $\sigma_{12}$ gives
\begin{align}
\sigma_{12}\,\sigma_{23}\,\sigma_{12}=\sigma_{13}\,,\qquad
\sigma_{12}\,\sigma_{123}\,\sigma_{12}=\sigma_{132}\,.
\end{align}
Therefore, for the $\sigma_{12}$ and $\sigma_{123}$-twisted Cardy states,
one simply has
\begin{align}
\sigma_{12}\ket{b_{\lambda}^{\sigma_{23}}}_B=\ket{b_{\sigma_{12}(\lambda)}^{\sigma_{13}}}_B\,,\qquad       \sigma_{12}\ket{b_{\lambda}^{\sigma_{123}}}_B=\ket{b_{\sigma_{12}(\lambda)}^{\sigma_{132}}}_B\,.
    \label{eq:conj_sigma}
\end{align}
Since $\sigma_{12}$ is invertible, one can run over all four $\sigma_{13}$-twisted or two $\sigma_{132}$-twisted Cardy states, respectively.

We now pull these states back to $A$. The associated topological interface $\mathcal I_{B\to A}$ is defined by gauging the $\sigma_{12}$-automorphism in the tetrality web in Fig.~\ref{fig:web}, i.e. $A=B^{\langle \sigma_{12}\rangle}$. The pair of dual interfaces $\mathcal I_{A\to B}$ and $\mathcal I_{B\to A}$ satisfy the following identities,
\begin{align}
 \cI_{B\to A}\circ\cI_{A\to B}
 &=\one_A+{j_A}\,,
 &\cI_{A\to B}\circ\cI_{B\to A}
 &=\one_B+\sigma_{12}\,,
 \label{eq:interfaces}
\end{align}
From Eq.~\eqref{eq:conj_sigma}, it is easy to see that all these $\sigma$-twisted Cardy states are \emph{not} $\sigma_{12}$-invariant. Therefore, according to~\cite{Fukusumi:2021zme}, the pulled-back states in $A$ must be $\sigma_{12}$-invariant, and thus have
to be the linear combinations of the $\mathbb Z_2^{\sigma_{12}}$-orbit, 
\begin{align}
 \cI_{B\to A}\ket{b^g_\lambda}_B
 =\cI_{B\to A}\ket{b^{\sigma_{12}g\,\sigma_{12}}_{\sigma_{12}(\lambda)}}_B
 =\frac{1}{\sqrt2}\left(\,\ket{b^g_\lambda}_B+\ket{b^{\sigma_{12}g\,\sigma_{12}}_{\sigma_{12}(\lambda)}}_B\right)\,,\quad{\rm for}\ \ g\in\mathcal S_\sigma\,.
 \label{eq:orbit}
\end{align}
Consequently, the two transpositions lead to the same $4$ physical boundaries, and the two 3-cycles give the same $2$ as well. Overall, we end up with $4+2=6$ non-factorizable boundaries after gauging back to the physical theory $A$. For each auxiliary state, we denote its physical pullback by
\begin{align}
 \ket{M^g_\lambda}_A
 =\mathcal I_{B\to A}\ket{b^g_\lambda}_B\,.
 \label{eq:BphysicalBoundaries}
\end{align}
The explicit expansion in terms of the Ishibashi states in theory $A$ follows from the
brane transport formula in~\cite{Chen:2026byx} that we omit here for brevity. Since every permutation fixes the chiral algebra $D=B^{S_3}$, these physical branes $\ket{M^g_\lambda}_A$ preserve ordinary $D$ gluing. Below, we will identify them with the six boundaries constructed in Section~\ref{sec:Dprojector}.

\subsection{Identifying boundaries from the two routes}
\label{sec:identifyBD}

The two brane constructions from either moving down or up approaches can be identified with each other from their mixed annulus partition functions. First, it is easy to show that the $6$ branes $\ket{M^g_\lambda}_A$ are simple. Equation~\eqref{eq:interfaces} expresses their annulus amplitudes as the sum of two overlaps in the auxiliary theory $B$,
\begin{align}
 Z^A_{M^g_\lambda,M^g_\lambda}
 =Z^B_{b^g_\lambda,b^g_\lambda}
  +Z^B_{b^g_\lambda,\sigma_{12}(b^g_\lambda)}\,.
 \label{eq:BimageSelfAnnulus}
\end{align}
The first term has one vacuum contribution as follows from Eq.~\eqref{eq:BtwistedCardyCondition}. The second term is an overlap between two distinct elementary boundaries exchanged by $\sigma_{12}$, and therefore contains no vacuum state. The physical self-annulus in Eq.~\eqref{eq:BimageSelfAnnulus} thus contains the vacuum exactly once, so each pulled back boundary is elementary. As explained in Section~\ref{sec:Bproject}, the permutation $\sigma\in\mathcal S_\sigma$ gluing mixes the two stress tensors and gives nonzero energy transmission. These simple boundaries are therefore distinct from every product boundary $\mathcal F_a$, so $n_0(\mathcal F_a,M^g_\lambda)=0$.

To compare them with the $D$ construction, we first calculate their mixed amplitudes with the factorized boundary $\mathcal F_{0,0}$, namely $\ket{\mathcal F_{0,0}}_A=\mathcal I_{D\to A}\ket0_D$, by acting with the interface $\mathcal I_{D\to A}$ on the $\ket{M^g_\lambda}_A$ side,
\begin{align}
 Z^A_{\mathcal F_{0,0},M^g_\lambda}(q)
 =\sum_i n_i^{g,\lambda}\chi_i(q)\,,\qquad
 \mathcal I_{A\to D}\ket{M^g_\lambda}_A
 =\sum_i n_i^{g,\lambda}\ket i_D\,,
 \label{eq:BimageReverse}
\end{align}
where $n_i^{g,\lambda}$ is the multiplicity of the open $D$ character $\chi_i$ in this annulus. On the other hand, we can also apply Eq.~\eqref{eq:BphysicalBoundaries} and move the topological interface $\mathcal I_{B\to A}$ to the $\mathcal F_{0,0}$ side, and thus have
\begin{align}
    Z^A_{\mathcal F_{0,0},M^g_\lambda}(q)=Z^B_{b_0,b_\lambda^g}(q)=\chi_\lambda^{{\rm tw},g}(q)\,,
\end{align}
where $b_0$ is the untwisted Cardy brane in $B$ with the identity label. Since $D$ is the smallest algebra obtained by the $S_3$ gauging of $B$, these twisted characters can be expanded in terms of $\chi_i$ as summarized in Table~\ref{tab:pf_FM}. We can now compare the physical boundaries in the $A$ theory themselves. Using $\ket{\mathcal B_i}_A=\mathcal I_{D\to A}\ket i_D$ and Eq.~\eqref{eq:BimageReverse}, the vacuum multiplicity is $n_0(\mathcal B_i,M^g_\lambda)=n_i^{g,\lambda}$. Removing the product components as in Eq.~\eqref{eq:Dremainder} then gives
\begin{align}
 n_0(\mathcal R_i,M^g_\lambda)
 &=n_i^{g,\lambda}
   -\sum_a b_{ai}\,n_0(\mathcal F_a,M^g_\lambda)
 =n_i^{g,\lambda}\,.
 \label{eq:crossConstructionVacuum}
\end{align}
Taking $i=\frac{1}{16},\frac{13}{80},\frac{7}{16}, \frac{3}{80},\frac{1}{9},\frac{2}{45}$ for the $6$ ordinary Cardy branes from the $D$ side, these coefficients $n_i^{g,\lambda}$ form the $6\times6$ identity matrix. Since both boundaries $\mathcal R_i$ and $M^g_{\lambda}$ are elementary, they have to be identical to each other. We also summarize their relations in Table~\ref{tab:pf_FM}. In particular, $\mathcal R_{\frac{1}{16}}$ is the pullback of $\ket{b^{\sigma_{23}}_{(1,1;1)}}_B$, which we identify with $\mathcal T_1$, Gaiotto's RG interface~\cite{Gaiotto:2012np}.  We will use both the $B$ and $D$ theory constructions to compute the action of topological lines and the boundary observables in the next section.

\begin{table}[t]
 \centering
 \begin{tabular}{c c c c}
 \toprule
 $\sigma$-twists  & $\lambda$-labels & twisted-$B$ characters & brane notations\\
 \midrule
 $\sigma_{23}$ & $(1,1;1)$ & $\chi_{\frac{1}{16}}+\chi_{\frac{25}{16}}$ & $\mathcal T_1=\mathcal R_{\frac{1}{16}}$\\[5pt]
 $\sigma_{23}$ & $(1,1;3)$ & $\chi_{\frac{13}{80}}+\chi_{\frac{53}{80}}$ & $\mathcal T_\tau=\mathcal R_{\frac{13}{80}}$\\[5pt]
 $\sigma_{23}$ & $(2,2;1)$ & $\chi_{\frac{7}{16}}+\chi_{\frac{15}{16}}$ & $\mathcal T_\eta=\mathcal R_{\frac{7}{16}}$ \\[5pt]
 $\sigma_{23}$ & $(2,2;3)$ & $\chi_{\frac{3}{80}}+\chi_{\frac{43}{80}}$ & $\mathcal T_{\eta\tau}=\mathcal R_{\frac{3}{80}}$\\[5pt]
 $\sigma_{123}$ & $(1;1)$ & $\chi_{\frac{1}{9}}+\chi_{\frac{7}{9}}+\chi_{\frac{13}{9}}$ & $\mathcal T_\sigma=\mathcal R_{\frac{1}{9}}$\\[5pt]
 $\sigma_{123}$ & $(1;3)$ & $\chi_{\frac{2}{45}}+\chi_{\frac{17}{45}}+\chi_{\frac{32}{45}}$ & $\mathcal T_{\sigma\tau}=\mathcal R_{\frac{2}{45}}$\\[5pt]
 \bottomrule
 \end{tabular}
 \caption{The mixed annulus partition functions between $\mathcal F_{0,0}$ and $M_\lambda^g$}
 \label{tab:pf_FM}
\end{table}

\section{Non-invertible symmetries and Interface spectra}
\label{sec:spectra}\label{sec:observables}

%\JC{sec 5 is done} \ALA{Went through it and made only minor language changes} \JR{all good}

Fusing a topological defect line (TDL) into one of the non-factorizable interfaces will generally produce a linear combination of them. This action relates all $6$ non-trivial boundaries and organizes their spectra. In this subsection, we use these TDLs to understand the annulus amplitudes and then compare the physical observables of the interfaces. Their energy transmission is the same (since the lines are topological) but their boundary entropies and 1d local operators can distinguish them.

\subsection{Topological lines acting on the interfaces}
\label{sec:lines}

From the moving up approach, it is clear that all $6$ interfaces preserve the diagonal
$\mathrm{TY}(\mathbb Z_2)$ fusion category shared by Ising and TIM. In Ising, we label their simple lines by $\{\one,\psi,\sigma'\}$. In TIM, the full fusion category is given by the tensor product ${\rm Fib}\boxtimes \mathrm{TY}(\mathbb Z_2)$, whose TDLs are labelled by $\{\one, \tau\}$ and $\{\one, \eta, \sigma\}$, respectively~\cite{Chang:2018iay,Thorngren:2021yso}. By preserving the fusion category we mean that the shared TDLs in $\mathrm{TY}(\mathbb Z_2)$ of Ising and TIM can pass through the interface, with $\psi$ becoming $\eta$ and $\sigma'$ becoming $\sigma$. In the folded theory $A$, this is equivalent to the fact that the tensor product  lines $\mathcal L_{\psi\,\boxtimes\,\eta}$ and $\mathcal L_{\sigma'\boxtimes\,\sigma}$ can end on the boundaries. Therefore, the boundary states obey
\begin{align}
 \cL_{\psi\otimes\one}\ket{\mathcal T_a}_A
 =\cL_{\one\otimes\eta}\ket{\mathcal T_a}_A\,,\qquad
 \cL_{\sigma'\otimes\one}\ket{\mathcal T_a}_A
 =\cL_{\one\otimes\sigma}\ket{\mathcal T_a}_A\,,
 \label{eq:diagonalTY}
\end{align}
for each of the $6$ boundary labels $a$. Thus, the action from the Ising side is already fixed by the action from the TIM side, which we now study.

The 6 Verlinde lines in TIM can be explicitly labeled by $\one,\tau,\eta,\eta\tau,\sigma,\sigma\tau$. To determine their actions, it is enough to start with the RG boundary $\mathcal T_1$. On a module $A_{(u,v)}$ in $A$, $\cL_{\one\otimes a}$ multiplies every Ishibashi state by $\frac{S^{\TIM}_{av}}{S^{\TIM}_{0v}}$~\cite{Petkova:2000ip}. Multiplying these factors to Eq.~\eqref{eq:T0Dprojectors}, we have
\begin{align}
 \cL_{\one\otimes a}\ket{\mathcal T_1}_A
 =\ket{\mathcal T_a}_A\,,
 \qquad
 a\in\mathcal S_T\equiv\{\one,\tau,\eta,\eta\tau,\sigma,\sigma\tau\}\,,
 \label{eq:TIMseedaction}
\end{align}
where the resulting states are precisely the six boundaries identified in Table~\ref{tab:pf_FM}, with $\mathcal T_{\one}\equiv \mathcal T_1$. This also justifies the subscripts we had previously introduced in Section~\ref{sec:Dprojector}.

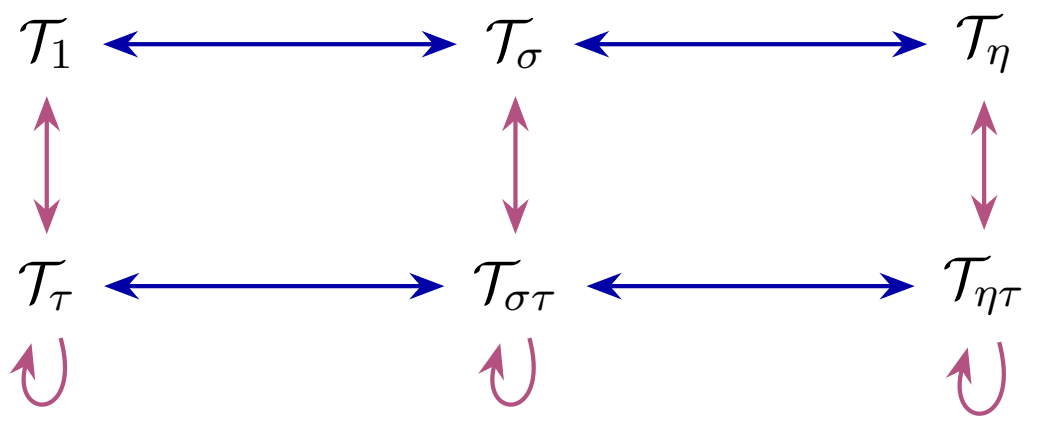
\begin{figure}[htbp]
\centering
\begin{tikzpicture}[>=Stealth,x=3.1cm,y=1.6cm,
 every node/.style={inner sep=5pt}]
\node (a) at (0,1) {$\mathcal T_1$};
\node (b) at (1,1) {$\mathcal T_\sigma$};
\node (c) at (2,1) {$\mathcal T_\eta$};
\node (af) at (0,0) {$\mathcal T_\tau$};
\node (bf) at (1,0) {$\mathcal T_{\sigma\tau}$};
\node (cf) at (2,0) {$\mathcal T_{\eta\tau}$};
\draw[<->,thick,blue!65!black] (a)--(b);
\draw[<->,thick,blue!65!black] (b)--(c);
\draw[<->,thick,blue!65!black] (af)--(bf);
\draw[<->,thick,blue!65!black] (bf)--(cf);
\draw[<->,thick,magenta!70!black] (a)--(af);
\draw[<->,thick,magenta!70!black] (b)--(bf);
\draw[<->,thick,magenta!70!black] (c)--(cf);
\draw[->,thick,magenta!70!black] (af) edge[loop below] (af);
\draw[->,thick,magenta!70!black] (bf) edge[loop below] (bf);
\draw[->,thick,magenta!70!black] (cf) edge[loop below] (cf);
\end{tikzpicture}
\caption{The TIM lines acting on the six interfaces. Blue arrows represent $\sigma$ and magenta arrows represent $\tau$.}
\label{fig:nim}
\end{figure}

The action on every other boundary easily follows by fusing the TIM lines. For example, the KW duality line $\sigma^2=\one+\eta$ gives
\begin{align}
 \cL_{\one\otimes\sigma}\ket{\mathcal T_\sigma}_A
 =\ket{\mathcal T_1}_A+\ket{\mathcal T_\eta}_A\,.
 \label{eq:dualitydecoration}
\end{align}
Similarly, $\eta$ exchanges $\mathcal T_1$ and $\mathcal T_\eta$ and fixes $\mathcal T_\sigma$, while $\tau$ sends $\mathcal T_a$ to $\mathcal T_{a\tau}$ and $\mathcal T_{a\tau}$ to $\mathcal T_a+\mathcal T_{a\tau}$ for $a=1,\eta,\sigma$. We summarize these relations in the quiver diagram in Figure~\ref{fig:nim}.

More generally, let $N_{ab}^c$ be the TIM fusion multiplicities, with $a,b,c$ running over its $6$ simple lines. Associativity of line fusion gives
\begin{align}
 \cL_{\one\otimes a}\ket{\mathcal T_b}_A
 =\cL_{\one\otimes a}\cdot\cL_{\one\otimes b}
       \ket{\mathcal T_1}_A
 =\sum_{c\in \mathcal S_T}N_{ab}^c\,\ket{\mathcal T_c}_A\,.
 \label{eq:regularTIM}
\end{align}
The 6 branes thereby carry the regular representation of the TIM fusion ring, known as a NIM-rep~\cite{Behrend:1999bn}, whose representation matrices are simply $(N_a)_b^c\equiv N_{ab}^c$. The three generating matrices are
\begin{align}
 N_\eta
 &=\begin{pmatrix}0&1&0\\1&0&0\\0&0&1\end{pmatrix}
      \otimes\one_2\,,
 &N_\sigma
 &=\begin{pmatrix}0&0&1\\0&0&1\\1&1&0\end{pmatrix}
      \otimes\one_2\,,
 &N_\tau
 &=\one_3\otimes\begin{pmatrix}0&1\\1&1\end{pmatrix}\,.
 \label{eq:generatinglineactions}
\end{align}
These matrices themselves also satisfy
\begin{align}
 N_a\cdot N_b
 =\sum_{c\in \mathcal S_T} N_{ab}^c\,N_c\,,
 \qquad N_{\one}=\one_6\,.
 \label{eq:NIMrelations}
\end{align}

\subsection{Annulus partition functions}

We will now determine the spectra on and between the six non-factorizable interfaces by computing their mixed annulus partition functions in the open string channel. In the folded picture, for a pair of boundary conditions $\mathcal T_a$ and $\mathcal T_b$, let $\mathcal H_{ab}$ be the Hilbert space on a strip with these conditions at its ends. The annulus partition functions in the open
and closed string channels are~\cite{Cardy:1989ir,Fuchs:2002cm}
\begin{align}
 Z^A_{a,b}(q)
 &=\tr_{\mathcal H_{ab}}q^{L_0-\frac{c}{24}}
 =_A\!\bra{\mathcal T_a} \widetilde q^{\frac{1}{2}(L_0+\bar L_0-\frac{c}{12})}\ket {\mathcal T_b}_{A}
 =\sum_{i\in\operatorname{Irr}(D)}(n_i)_{ab}\,\chi_i(q)\,,
 \label{eq:annulus}
\end{align}
For $a=b$ this is the spectrum of local interface operators, whereas for distinct boundaries it is the spectrum of boundary condition changing fields.

To compute the $6$ amplitudes with $\mathcal T_1$ at one end, we use the decomposition in Eq.~\eqref{eq:T0Dprojectors}. For $a\in\mathcal S_T$, we therefore have
\begin{align}
 Z^A_{1,a}(q)
 =Z^D_{\frac{1}{16},\,\mathcal I_{A\to D}\mathcal T_a}(q)
   -Z^A_{\mathcal F_{\frac{1}{16},0},\,\mathcal T_a}(q)\,,
 \label{eq:annulusfromtransport}
\end{align}
The 6 transported states are already given in Eq.~\eqref{eq:sixrestrictions}. Each term in the first amplitude is thus an ordinary Cardy annulus amplitude $Z^D_{ij}(q)=\sum_kN^D_{ij}{}^k\chi_k(q)$, where $N^D_{ij}{}^k$ are the fusion coefficients of $D$. The second amplitude follows from the shared duality line in Eq.~\eqref{eq:diagonalTY}. The product boundary $\mathcal F_{\frac{1}{16},0}$ is obtained by fusing the Ising line $\sigma'$ onto $\mathcal F_{0,0}$. Moving it onto $\mathcal T_a$ turns it into the TIM line $\sigma$, whose action has been given in Eq.~\eqref{eq:regularTIM}. Since $|\mathcal F_{0,0}\rangle_A=\mathcal I_{D\to A}|0\rangle_D$, this gives
\begin{align}
 Z^A_{\mathcal F_{\frac{1}{16},0},\,\mathcal T_a}(q)
 =\sum_cN_{\sigma a}^{c}\,
            Z^A_{\mathcal F_{0,0},\,\mathcal T_c}(q)
 =\sum_cN_{\sigma a}^{c}\,
            Z^D_{0,\,\mathcal I_{A\to D}\mathcal T_c}(q)\,.
 \label{eq:annulusproductsubtraction}
\end{align}
The subtraction is therefore read directly from the same six sums in Eq.~\eqref{eq:sixrestrictions}. For example, taking $a=1$, we obtain
\begin{align}
 Z^A_{1,1}
 &=Z^D_{\frac{1}{16},\frac{1}{16}}
   +Z^D_{\frac{1}{16},\frac{25}{16}}
   -\left(\chi_{\frac{1}{9}}+\chi_{\frac{7}{9}}
                 +\chi_{\frac{13}{9}}\right)\,,
 \label{eq:annulusselftransport}
\end{align}
The $6$ mixed annulus partition functions between $\mathcal T_1$ and $\mathcal T_a$ can be similarly computed as
\begin{align}
 &Z^A_{1,1}
 =\chi_0+\chi_3+2\chi_2
   +\chi_{\frac12}+\chi_{\frac32}
   +\chi_{\frac19}+\chi_{\frac79}+\chi_{\frac{13}9}\,,\notag\\[5pt]
 &Z^A_{1,\tau}
 =\chi_{\frac{13}5}+\chi_{\frac85}+2\chi_{\frac35}
   +\chi_{\frac{11}{10}}+\chi_{\frac1{10}}
   +\chi_{\frac2{45}}+\chi_{\frac{17}{45}}+\chi_{\frac{32}{45}}\,,\notag\\[5pt]
& Z^A_{1,\eta}
 =2\chi_{\frac12}+2\chi_{\frac32}
   +\chi_{\frac19}+\chi_{\frac79}+\chi_{\frac{13}9}\,,\qquad
 Z^A_{1,\mathcal \eta\tau}
 =2\chi_{\frac{11}{10}}+2\chi_{\frac1{10}}
   +\chi_{\frac2{45}}+\chi_{\frac{17}{45}}+\chi_{\frac{32}{45}}\,,\notag\\[5pt]
& Z^A_{1,\sigma}=2\left(\chi_{\frac1{16}}+\chi_{\frac{25}{16}}
                  +\chi_{\frac7{16}}+\chi_{\frac{15}{16}}\right)\,,\qquad
 \ \ \, Z^A_{1,\mathcal {\sigma\tau}}
 =2\left(\chi_{\frac{53}{80}}+\chi_{\frac{13}{80}}
                  +\chi_{\frac3{80}}+\chi_{\frac{43}{80}}\right)\,,
 \label{eq:annulusseedssigma}
\end{align}
where only the first amplitude contains the identity, as expected for $6$ distinct elementary boundaries.

All other annulus amplitudes now follow from the regular TDL actions in Figure~\ref{fig:nim}. Since every TIM line is self dual, the line creating $\mathcal T_a$ from $\mathcal T_1$ can be moved onto the other boundary. Using Eq.~\eqref{eq:regularTIM}, we obtain
\begin{align}
 Z^A_{a,b}(q)
 =\sum_c N_{ab}^{c}\,
 Z^A_{1,c}(q)\,.
 \label{eq:annulusTIMfusion}
\end{align}
For example, the fusion rules $\sigma^2=1+\eta$ and $\tau^2=1+\tau$ give
\begin{align}
 Z^A_{\sigma,\sigma}
 &=Z^A_{1,1}
   +Z^A_{1,\mathcal \eta}\,,&
 Z^A_{\tau,\tau}
 &=Z^A_{1,1}
   +Z^A_{1,\tau}\,.
 \label{eq:annulusselfexamples}
\end{align}
The $6$ amplitudes in Eq.~\eqref{eq:annulusseedssigma} above therefore determine all $36$ amplitudes.

\subsection{Boundary entropy and energy transmission}

We now compute the boundary entropy and energy reflection and transmission coefficients of the six non-factorizable interfaces. For a boundary state $\ket {\mathcal T_a}_A$, the boundary entropy and the associated $g$-factor is given by~\cite{Affleck:1991tk}
\begin{align}
    S_a=\log g_a\,,\qquad
    g_a\equiv{}_A\langle0|\mathcal T_a\rangle_A\,.
\end{align}
Since the topological interface is defined by a $\mathbb Z_2$-orbifolding, it has vacuum amplitude $ {}_A\langle0|\cI_{B\to A}=\sqrt2\,{}_B\langle0|$. Therefore only the vacuum Ishibashi state in Eq.~\eqref{eq:seed} contributes to this overlap. Overall we have
\begin{align}
 g_{M^g_\lambda}
 =\sqrt2\,\frac{\Psi^g_{\lambda0}}{\sqrt{S^B_{00}}}\,,
 \label{eq:gfromvacuum}
\end{align}
where the relation between labels $a\in\mathcal S_T$ and $(\sigma,\lambda)$ are encoded in Table~\ref{tab:pf_FM}. Substituting the explicit states of Eq.~\eqref{eq:referenceSeeds} gives
\begin{align}
 g_{M^{\sigma_{23}}_{(b,b;d)}}
 &=2^{\frac{3}{4}}\frac{S^{(3)}_{d1}}{\sqrt{S^{(3)}_{11}}}\,,
 &g_{M^{\sigma_{123}}_{(1;d)}}
 &=2^{\frac{5}{4}}\frac{S^{(3)}_{d1}}{\sqrt{S^{(3)}_{11}}}\,,
 \qquad b=1,2\,,\quad d=1,3\,.
 \label{eq:gseeds}
\end{align}
Writing $g_0=\left(2-\frac2{\sqrt5}\right)^{\frac{1}{4}}$ and $\fphi=\frac{(1+\sqrt5)}2$, the six
vacuum overlaps are
\begin{align}
 (g_{1},g_{\tau},g_{\eta},
  g_{{\eta\tau}},g_{\sigma},
  g_{{\sigma\tau}})
 =g_0(1,\fphi,1,\fphi,\sqrt2,\sqrt2\fphi)\,.
 \label{eq:gsix}
\end{align}
As a consistency check, note that all $6$ interfaces are connected with each other by fusing different TDLs in $\mathcal S_T$. Since the $g$-factors of these TDLs are their quantum dimensions, we find that the relative ratios of the $g$-factors in Eq.~\eqref{eq:gsix} match perfectly with Eq.~\eqref{eq:regularTIM}.

On the other hand, energy reflection and transmission are determined by the normalized stress tensor overlaps~\cite{Quella:2006de}. With $i,j\in\{\Is,\TIM\}$ and $c_{\Is}=\frac12$, $c_{\TIM}=\frac{7}{10}$, we define the stress-tensor overlaps $R_{ij}$ and the transmission ($\mathsf T$) and reflection ($\mathsf R$) coefficients
\begin{align}
 \label{eq:energydefinition}
 &R_{ij}(M_\lambda^g)
 =\frac{{}_A\langle0|T^{(i)}(\infty)\, {\bar T^{(j)}}(0)|M_\lambda^g\rangle_A}{{}_A\langle0|M_\lambda^g\rangle_A}
 ={}_A\left \langle0\right|T^{(i)}(\infty)\, g\left({\bar T^{(j)}}(0)\right)\left|0\right\rangle_A
 \,,\\[5pt]
 &\mathsf T_M
 =\frac{2(R_{\Is,\TIM}+R_{\TIM,\Is})}{c_{\Is}+c_{\TIM}}\,,\qquad
 \mathsf R_M
 =\frac{2(R_{\Is,\Is}+R_{\TIM,\TIM})}{c_{\Is}+c_{\TIM}}\,,\notag
\end{align}
where we have moved the topological interface $\cI_{A\to B}$ onto the stress tensor insertions and used the gluing automorphism $g$ in $B$. Since the stress tensors also belong to the vacuum module of $B$, the $\lambda$-dependent boundary vacuum amplitude factors out of the numerator and cancels against the denominator. The transmission and reflection rate therefore depend only on the gluing automorphism $g$. This relates energy transport to the mixing of the stress tensors with phantom currents~\cite{Furuta:2025ahl}. We use the ordered basis $(T_{\Is},T_{\TIM},X)$ of the spin-two subspace in Eq.~\eqref{eq:BvacuumMixing}, with $\langle X(z)X(0)\rangle=z^{-4}$. As discussed in~\cite{Gaiotto:2012np,Chen:2026byx}, the automorphism $g$ induces a matrix representation $\Omega_2(g)$ on the spin-two subspace,
\begin{align}
 g\begin{pmatrix}T_{\Is}\\T_{\TIM}\\X\end{pmatrix}
 =\Omega_2(g)\begin{pmatrix}T_{\Is}\\T_{\TIM}\\X\end{pmatrix}\,.
 \label{eq:stressbasis}
\end{align}
The four permutations give
\begin{align}
 \Omega_2(\sigma_{23})
 &=\frac18\begin{pmatrix}
 1&5&\sqrt7\\
 7&3&-\sqrt7\\
 4\sqrt7&-\frac{20}{\sqrt7}&4
 \end{pmatrix}\,,
 &\Omega_2(\sigma_{13})
 &=\frac18\begin{pmatrix}
 1&5&-\sqrt7\\
 7&3&\sqrt7\\
 -4\sqrt7&\frac{20}{\sqrt7}&4
 \end{pmatrix}\,,
 \notag\\[6pt]
 \Omega_2(\sigma_{123})
 &=\frac18\begin{pmatrix}
 1&5&-\sqrt7\\
 7&3&\sqrt7\\
 4\sqrt7&-\frac{20}{\sqrt7}&-4
 \end{pmatrix}\,,
 &\Omega_2(\sigma_{132})
 &=\frac18\begin{pmatrix}
 1&5&\sqrt7\\
 7&3&-\sqrt7\\
 -4\sqrt7&\frac{20}{\sqrt7}&-4
 \end{pmatrix}\,.
 \label{eq:M2cycles}
\end{align}
Notice that the two by two stress tensor blocks are identical in all four matrices. Applying Eq.~\eqref{eq:energydefinition}, we find, for each of the six interfaces,
\begin{align}
 R(M_\lambda^g)&=\frac1{160}\begin{pmatrix}5&35\\35&21\end{pmatrix}\,,
 &\mathsf T_M&=\frac{35}{48}\,,
 &\mathsf R_M&=\frac{13}{48}\,.
 \label{eq:transmission}
\end{align}
The result is also consistent with Eq.~\eqref{eq:regularTIM}. Since the attached topological interfaces are completely transparent and have transmission rate $\mathsf T=1$, the $6$ conformal interfaces must have the same transmission and reflection rates as we found in \eqref{eq:transmission}.

\section{Discussion}\label{sec:discussion}

We have shown that the Ising and tricritical Ising theories admit six
transmitting interfaces preserving the common rational
chiral algebra $D$. They can be reached in two ways.
Moving up exposes permutations in the larger coset $B$,
which mix the physical stress tensors and specify
twisted Cardy states. Moving down keeps the currents
shared by these gluing conditions and constructs
ordinary Cardy states of $D$. Transport through the
corresponding topological interfaces brings both
descriptions back to the same physical theory.

The tetrality web explains the relation between the
two procedures. The physical product theory $A$ is fixed by a
transposition in $B$, and the common algebra $D$ is fixed
by the whole group $S_3$. The moving down construction
gives twenty four elementary boundaries, of which
eighteen are factorized and six transmit energy.
The moving up construction supplies explicit states
for the latter: four come from a transposition and
two from a 3-cycle. The current of spin six in $D$
has a nonzero self-coupling, which fixes its action
under a conformal automorphism and leaves ordinary
gluing as the common boundary condition.

The topological lines organize the six transmitting interfaces
into the regular fusion representation of TIM.
In particular, the 3-cycle interface is obtained by
fusing a duality line into the Gaiotto wall.
A topological line multiplies the vacuum and its
stress-tensor descendants by the same eigenvalue,
so these operations preserve normalized energy
transmission. This gives a physical explanation
for the common value $\frac{35}{48}$. Other bulk sectors
respond differently to the line. Their overlaps,
together with the boundary entropies and the open
spectra, resolve the differences among the interfaces.

The explicit annulus amplitudes that we obtained make boundary RG flows a natural
next question. Their spectra identify the relevant fields
on each interface and the operators that change one
interface into another. Computing their boundary
operator products would determine how the six
transmitting interfaces and the factorized boundaries
are connected by perturbations, at least to leading order in conformal perturbation theory. On the lattice,
the same question can be addressed by varying local
couplings at an Ising/TIM junction~\cite{Cogburn:2023xzw,Antunes:2026kqb}. The exact
spectra and correlations derived here give concrete
quantities with which to identify the resulting
interfaces. Furthermore, the $g$-theorem gives an organizing principle for this complicated network of RG flows~\cite{Affleck:1991tk,Friedan:2003yc}.

Moving down also raises the question of which currents
must survive at a rational interface. In this example,
the OPE approach excludes a smaller strongly
rational conformal algebra whose first extra current
has weight below twenty five. It would be interesting
to continue this analysis and to apply the same
physical construction to other coset models.
The combination of an auxiliary extension and a
common rational subalgebra gives complementary
ways to organize interfaces, calculate their
states and understand how energy and operators
pass between critical systems.

Finally, while we have conjectured a classification of rational boundaries for $c=\frac{6}{5}$ this does not exhaust the space of conformal boundary conditions. Irrational boundaries, in particular ones preserving only the total Virasoro stress-tensor, are of great interest as they form an interesting bridge between the realms of solvable and unsolvable 2d CFTs. Some tools to understand this problem are already available: In \cite{Chen:2026byx}, the author used a primal modular bootstrap approach to study similar interfaces in the tetracritical Ising to 3-state Potts RG domain wall, and the dual bootstrap of \cite{Meineri:2025qwz} is sufficiently versatile to derive bounds in the space of irrational boundary conditions.
In addition, the Kormos-Runkel-Watts interfaces~\cite{Kormos:2009sk}, especially the conformal defect for the $M(5,6)$ minimal model, which was recently studied using lattice-model construction in~\cite{Rong:2026xxg}, have eluded any rational conformal interface construction. 
It will be interesting to run our search algorithm to look for possible analytical understanding.

\paragraph{Acknowledgements}
We thank Costas Bachas and Yunqin Zheng for useful discussions and Bufen Zhang for coordinating submission. AA further thanks Andreia Gon\c{c}alves for continued inspiration. AA is funded by the European Union (ERC, FUNBOOTS, project number 101043588, PI Miguel Paulos). JR is funded by the European Union (ERC, QFTinAdS, project number 101087025, PI Balt van Rees). JC is supported by the Fujian Provincial Natural Science Foundation of China (No. 2025J01004), and the National Natural Science Foundation of China (Grant No. 12247103). Views and opinions expressed are however those of the authors only and do not necessarily reflect those of the European Union or the European Research Council Executive Agency. Neither the European Union nor the granting authority can be held responsible for them.

\paragraph{AI Disclosure}
We have used LLMs (Claude, ChatGPT) to develop the algorithmic search for rational subalgebras of $D$, which found no candidates up to weight 25. An illustrative example is given in Appendix \ref{app:opeconstraints}, which is also mostly written by LLMs.

\appendix
\section{OPE constraints on rational conformal subalgebras}
\label{app:opeconstraints}

To look for interfaces that preserve fewer chiral currents, we
search for rational conformal subalgebras of $B$ with the same
stress tensor.
For such a subalgebra $U$, let $h_*(U)$ be the
lowest weight of a Virasoro primary beyond the identity in its
vacuum module. 
The self OPE of a current at this weight cannot
produce a primary at any lower positive weight. 
As described in Section~\ref{sec:methods}, imposing these conditions on a
general linear combination of primaries gives quadratic equations
for its coefficients.

Below, we check to see whether $h_*(U)=21$ is allowed for some $U\subseteq B$. 
At weight 21, we find the 3342 states of $B$; among them, there are just seven Virasoro primaries, denoted by $p_1,\ldots,p_7$. 
Consider a state $v=\sum_{i=1}^7 x_i p_i$ with arbitrary $x_i\in\mathbb C$. 

\textbf{From the OPE to eleven equations.} Consider the coefficient of $z^{k-42}$ in $v(z)v(0)$. 
The operator has weight $k$. If $0<k<21$, it can contain descendants of the vacuum, but the coefficients of every weight-$k$ primary $a$ must vanish: otherwise the OPE has produced a primary component below $h_*(U)$.
Let $M^{(a)}_{ij}$ be the coefficients for $p_i(z)p_j(0)$ OPE. 
The condition on $v$ is the quadratic equation
\[
 q_a(x)=\sum_{i,j=1}^7 M^{(a)}_{ij}x_i x_j=0 .
\]
We use 11 such equations: two from $a$ at weight 2, two from weight 4, one from weight 6, three from weight 8, and three from weight 10. 
We will see that these equations lead to $x_i=0$ for all $x_i$'s.

\textbf{From eleven equations to one matrix.} The OPE coefficients in this basis are rational. 
For each equation $q_a$, multiply the whole equation by a common denominator of its coefficients. 
This gives polynomials $Q_a$ with integer coefficients.

If $Q_a(x)=0$, then $x_r x_s Q_a(x)=0$ for every $1\le r\le s\le7$. Multiply the equations by all possible $x_r x_s$; we get 308 degree-four equations.
There are however, only $\frac{7\cdot8\cdot9\cdot19}{4!}=210$ degree-four monomials in total.
The coefficients of these monomials form the so-called degree-four Macaulay matrix~\cite{MacAulay_1902,Lazard_1983}, which we denote as $M$.
If we denote $m_4(x)$ as a length 210 vector given by the 210 degree-four monomials, the equations can be written as
$$M \cdot m_4(x)=0.$$
For the equations to have non-zero solutions, all $210\times210$ sub-matrices (denoted by $N$) of $M$ must have zero determinant. 
The determinant is usually called a minor.
For an integer matrix \(M\), we can test its minors modulo a prime \(q\), working in the field \(\mathbb F_q=\mathbb Z/q\mathbb Z\)~\cite{DumasPernet2013}.
Our AI-generated algorithm found a matrix with 
\[
 \det N\equiv 611\pmod{1009}.
\]
This means that the equation holds if and only if $x_i=0$ for all $ x_i$. We therefore conclude that $h_*(U)=21$ is excluded. 

The above test needs an efficient way to calculate the OPE coefficients of theory $B$. 
We use the even part of a free Majorana fermion $\psi$ tensored with super-TIM, whose supercurrent is $G$.
Their product $\psi G$ is the weight-two field that extends $A$.
We compute composite OPEs recursively from the fermion anticommutator and the $c=7/10$ Neveu--Schwarz mode relations, combining the two factors with the fermionic signs. 
We also need to form a basis for the primaries. 
At each weight $h$ we list ordered products of creation modes of $\psi$, $G$ and $T_{\rm TIM}$ acting on the vacuum, retaining states of even total fermion parity. 
We remove the kernel of their Gram matrix and solve $L_1v=L_2v=0$ in the quotient using $L_n=L_n^{\rm Is}+L_n^{\rm TIM}$. 
Each independent solution gives a primary field.

\section{Chiral algebras, branching rules and modular data}
\label{app:modular}

\subsection{The representations of \texorpdfstring{$D$}{D}}

The interface spectra are sums of $D$ characters. We give the
representations, their modular transformations and their branching
from $A$, $B$ and $C$, so that the annulus amplitudes can be expanded
directly. We denote the irreducible $D$ representation of lowest
conformal weight $h$ by $D_h$ and define its character by
\begin{align}
 \chi_h(\tau)=\tr_{D_h}q^{L_0-\frac{1}{20}},\qquad
 q=e^{2\pi i\tau},\qquad \operatorname{Im}\tau>0\,.
 \label{eq:Dcharacter}
\end{align}
Each of the twenty four representations has one highest weight state, and
these all have different weights. Fusion with
$f=D_{\frac{13}{5}}$ pairs them in two sets of twelve as shown in Table~\ref{tab:Dmodules}.
The fusion rule $f\otimes f=\one\oplus f$ makes this pairing useful
for constructing the Fibonacci partners of the interfaces.

\begin{table}[htbp]
\centering
\begin{tabular}{c c c c}
\toprule
$x$ & $h(x)$ & $h(x\otimes f)$ & $\qdim(x)$\\
\midrule
$\one$ & $0$ & $\frac{13}{5}$ & $1$\\
$z$ & $3$ & $\frac{8}{5}$ & $1$\\
$J$ & $2$ & $\frac{3}{5}$ & $2$\\
$\rho_0$ & $\frac{1}{9}$ & $\frac{32}{45}$ & $4$\\
$\rho_1$ & $\frac{13}{9}$ & $\frac{2}{45}$ & $4$\\
$\rho_2$ & $\frac{7}{9}$ & $\frac{17}{45}$ & $4$\\
$\psi$ & $\frac{1}{2}$ & $\frac{11}{10}$ & $3$\\
$z\psi$ & $\frac{3}{2}$ & $\frac{1}{10}$ & $3$\\
$\Sigma_0$ & $\frac{1}{16}$ & $\frac{53}{80}$ & $3\sqrt2$\\
$z\Sigma_0$ & $\frac{25}{16}$ & $\frac{13}{80}$ & $3\sqrt2$\\
$\Sigma_1$ & $\frac{7}{16}$ & $\frac{3}{80}$ & $3\sqrt2$\\
$z\Sigma_1$ & $\frac{15}{16}$ & $\frac{43}{80}$ & $3\sqrt2$\\
\bottomrule
\end{tabular}
\caption{The $D$ representations grouped into Fibonacci pairs.
The first column labels a representation $x$, and the third gives the
lowest weight of $x\otimes f$. Its quantum dimension is
$\fphi\,\qdim(x)$, where $\fphi=\frac{1+\sqrt5}{2}$.
We order the modular matrices by the rows of this table, placing
$x$ and $x\otimes f$ next to each other.}
\label{tab:Dmodules}
\end{table}

With this ordering, the Fibonacci pairing separates the modular $S$
matrix into a twelve dimensional part and the Fibonacci matrix while the $T$ matrix is of diagonal form
\begin{align}
 S^D&=S^{\mathcal E}\otimes
 \frac{1}{\sqrt{\fphi+2}}
 \begin{pmatrix}1&\fphi\\ \fphi&-1\end{pmatrix},&
 T^D_{ii}&=e^{2\pi i(h_i-\frac{1}{20})}\,,
 \label{eq:SD}
\end{align}
where $\mathcal E$ labels the twelve rows in the first column of
Table~\ref{tab:Dmodules}. To write the twelve dimensional factor
explicitly, we use
\begin{align}
 a_9=-8\cos\frac{4\pi}{9},\qquad
 b_9=-8\cos\frac{8\pi}{9},\qquad
 c_9=-8\cos\frac{16\pi}{9},\qquad d_9=3\sqrt2\,.
\end{align}
The modular transformation on these twelve labels is then
\begingroup
\setlength{\arraycolsep}{3pt}
\begin{align}
S^{\mathcal E}=\frac1{12}
\begin{pmatrix}
1&1&2&4&4&4&3&3&d_9&d_9&d_9&d_9\\
1&1&2&4&4&4&3&3&-d_9&-d_9&-d_9&-d_9\\
2&2&4&-4&-4&-4&6&6&0&0&0&0\\
4&4&-4&a_9&b_9&c_9&0&0&0&0&0&0\\
4&4&-4&b_9&c_9&a_9&0&0&0&0&0&0\\
4&4&-4&c_9&a_9&b_9&0&0&0&0&0&0\\
3&3&6&0&0&0&-3&-3&-d_9&-d_9&d_9&d_9\\
3&3&6&0&0&0&-3&-3&d_9&d_9&-d_9&-d_9\\
d_9&-d_9&0&0&0&0&-d_9&d_9&0&0&6&-6\\
d_9&-d_9&0&0&0&0&-d_9&d_9&0&0&-6&6\\
d_9&-d_9&0&0&0&0&d_9&-d_9&6&-6&0&0\\
d_9&-d_9&0&0&0&0&d_9&-d_9&-6&6&0&0
\end{pmatrix}\,.
\label{eq:SE}
\end{align}
\endgroup
The fusion coefficients follow from the Verlinde formula~\cite{Verlinde:1988sn},
\begin{align}
 (N_i^D)_{jk}
 =\sum_m\frac{S^D_{im}S^D_{jm}(S^D_{km})^*}{S^D_{0m}}\,,
 \label{eq:Verlinde}
\end{align}
where $i,j,k,m$ run over the twenty four representations and $0$
is the vacuum label. Every representation is self conjugate, so the
fusion matrices $N_i^D$ are real and symmetric.

\subsection{Branching to the smallest algebra $D$}

To express a physical Ising--TIM observable in terms of the boundary-preserved
algebra $D$, we restrict each product representation $A_{x,y}$ as in
Table~\ref{tab:branching}. Here $x$ and $y$ are the Ising and TIM
weights. The table gives the rows of the $18\times24$ branching
matrix $b$ in Eq.~\eqref{eq:branch}, with the columns ordered by
Table~\ref{tab:Dmodules}. We use $x=0,\frac{1}{16},\frac{1}{2}$ in Ising and $x$,
$y=0,\frac{1}{10},\frac{3}{5},\frac{3}{2},\frac{3}{80},\frac{7}{16}$ in TIM.

\begin{table}[tbp]
\centering
\begin{tabular}{c c l}
\toprule
Ising weight $x$ & TIM weight $y$ & $\Res_D A_{x,y}$\\
\midrule
$0$ & $0$ & $D_0\oplus D_2$\\
$0$ & $\frac{1}{10}$ & $D_{\frac{1}{10}}$\\
$0$ & $\frac{3}{5}$ & $D_{\frac{3}{5}}\oplus D_{\frac{13}{5}}$\\
$0$ & $\frac{3}{2}$ & $D_{\frac{3}{2}}$\\
$0$ & $\frac{3}{80}$ & $D_{\frac{3}{80}}$\\
$0$ & $\frac{7}{16}$ & $D_{\frac{7}{16}}$\\
\midrule
$\frac{1}{16}$ & $0$ & $D_{\frac{1}{16}}$\\
$\frac{1}{16}$ & $\frac{1}{10}$ & $D_{\frac{13}{80}}$\\
$\frac{1}{16}$ & $\frac{3}{5}$ & $D_{\frac{53}{80}}$\\
$\frac{1}{16}$ & $\frac{3}{2}$ & $D_{\frac{25}{16}}$\\
$\frac{1}{16}$ & $\frac{3}{80}$ & $D_{\frac{1}{10}}\oplus D_{\frac{11}{10}}$\\
$\frac{1}{16}$ & $\frac{7}{16}$ & $D_{\frac{1}{2}}\oplus D_{\frac{3}{2}}$\\
\midrule
$\frac{1}{2}$ & $0$ & $D_{\frac{1}{2}}$\\
$\frac{1}{2}$ & $\frac{1}{10}$ & $D_{\frac{3}{5}}\oplus D_{\frac{8}{5}}$\\
$\frac{1}{2}$ & $\frac{3}{5}$ & $D_{\frac{11}{10}}$\\
$\frac{1}{2}$ & $\frac{3}{2}$ & $D_2\oplus D_3$\\
$\frac{1}{2}$ & $\frac{3}{80}$ & $D_{\frac{43}{80}}$\\
$\frac{1}{2}$ & $\frac{7}{16}$ & $D_{\frac{15}{16}}$\\
\bottomrule
\end{tabular}
\caption{The $D$ representations contained in each Ising--TIM product
representation $A_{x,y}$.}
\label{tab:branching}
\end{table}

We can also restrict the auxiliary $B$ states used in the
permutation-gluing construction. Only four sums of $D$ representations
occur,
\begin{align}
 U_B&=D_0\oplus D_3\oplus2D_2,&
 V_B&=D_{\frac{1}{2}}\oplus D_{\frac{3}{2}}\,,\notag\\
 U_{BF}&=D_{\frac{13}{5}}\oplus D_{\frac{8}{5}}\oplus2D_{\frac{3}{5}},&
 V_{BF}&=D_{\frac{11}{10}}\oplus D_{\frac{1}{10}}\,.
\end{align}
The eight $B$ representations, labelled as in
Eq.~\eqref{eq:Blabels}, decompose as
\begin{align}
\begin{array}{c|cccccccc}
\mu &[111;1]&[122;1]&[112;2]&[121;2]
     &[111;3]&[122;3]&[112;4]&[121;4]\\ \hline
\Res_D B_\mu&U_B&V_B&V_{BF}&V_{BF}&U_{BF}&V_{BF}&V_B&V_B
\end{array}\,.
\label{eq:BtoDbranch}
\end{align}
The repeated entries come from $B$ representations related by
numerator permutations and have the same
character decomposition in terms of the algebra $D$.

Extending $D$ by $\one\oplus z$ gives $C$. Four of its
representations combine pairs of $D$ modules,
\begin{align}
 D_0\oplus D_3,\qquad D_{\frac{13}{5}}\oplus D_{\frac{8}{5}},\qquad
 D_{\frac{1}{2}}\oplus D_{\frac{3}{2}},\qquad D_{\frac{11}{10}}\oplus D_{\frac{1}{10}}\,.
 \label{eq:Cpairedbranch}
\end{align}
The remaining sixteen representations form eight charge
conjugate pairs. The two members of each pair have the same
restriction to $D$, with lowest weight
\begin{align}
 2,\quad\frac35,\quad\frac19,\quad\frac{13}{9},\quad\frac79,
 \quad\frac{2}{45},\quad\frac{17}{45},\quad\frac{32}{45}\,.
 \label{eq:Cfixedbranch}
\end{align}
These two kinds of restriction give all twenty $C$ representations
and identify the ordinary and permutation sectors of $B$ inside the
intermediate $W_3$ theory.

\subsection{Characters}
\label{app:characters}

We now resolve the states in each $D$ representation. The character formulas below determine the conformal weights and descendant multiplicities appearing in the annulus amplitudes. Write $I_x$, $T_y$ and $P_z$ for the Virasoro characters of Ising, TIM and $\mathcal M_{5,6}$ (tetracritical Ising), with subscripts denoting their lowest weights.

The unexpected $\mathcal M_{5,6}$ characters enter through the lattice realization
of $B$ in~\cite{Dong_2004}. The chiral theory associated with the
rank two lattice $\sqrt{2}A_2$ has central charge $2$ and contains
commuting copies of $B$ and the three state Potts chiral algebra, with
central charges $\frac{6}{5}$ and $\frac{4}{5}$, respectively. Their
stress tensors add to the lattice stress tensor. The Virasoro
constituents of the Potts modules have characters $P_z$, with
$P_0+P_3$ giving the Potts vacuum character. The three cycle
$\sigma_{123}$ fixes this Potts algebra. Decomposing the lattice
modules twisted by this symmetry into twisted $B$ modules and ordinary
Potts modules therefore expresses the unknown $B$ characters in terms
of the known $P_z$. Together with their modular transformations, these
branching relations give the three cycle traces and twisted characters
used below~\cite[Proposition~6.1 and Theorem~6.2]{Dong_2004}.

The Virasoro characters $I_x$, $T_y$ and $P_z$ are
obtained from the minimal model formula~\cite{Rocha_Caridi_1985}
\begin{align}
\chi_{r,s}^{(p,p+1)}(\tau)
 =\frac1{\eta(\tau)}\sum_{n\in\mathbb Z}
 \left(
 q^{\frac{[2p(p+1)n+(p+1)r-ps]^2}{4p(p+1)}}
-q^{\frac{[2p(p+1)n+(p+1)r+ps]^2}{4p(p+1)}}
 \right)\,,
 \label{eq:Rocha}
\end{align}
where $\eta(\tau)=q^{\frac{1}{24}}\prod_{m\geq1}(1-q^m)$ is the
Dedekind eta function. The labels range over
$1\leq r\leq p-1$ and $1\leq s\leq p$, with the identification
$(r,s)\sim(p-r,p+1-s)$ and lowest weight
$h_{r,s}=\frac{[(p+1)r-ps]^2-1}{4p(p+1)}$.
Choosing $p=3,4,5$ gives $I,T,P$, respectively.

The two $B$ representations fixed by $S_3$ each split into three distinct
$D$ representations. To separate them, we insert the identity,
a transposition and a 3-cycle in the character trace. The resulting
functions are
\begin{align}
 F_0&=I_0T_0+I_{\frac{1}{2}}T_{\frac{3}{2}},&
 G_0&=I_0T_0-I_{\frac{1}{2}}T_{\frac{3}{2}},&
 R_0&=\frac{\eta(\tau)}{\eta(3\tau)}P_{\frac{1}{15}}\,,\notag\\
 F_1&=I_0T_{\frac{3}{5}}+I_{\frac{1}{2}}T_{\frac{1}{10}},&
 G_1&=I_0T_{\frac{3}{5}}-I_{\frac{1}{2}}T_{\frac{1}{10}},&
 R_1&=-\frac{\eta(\tau)}{\eta(3\tau)}P_{\frac{2}{3}}\,,
 \label{eq:stabletraces}
\end{align}
where the subscripts $0,1$ denote $[111;1]$ and $[111;3]$.
For the transposition trace $G_1$, we choose the action on
$I_0T_{\frac{3}{5}}$ to be positive. The projectors onto the trivial, sign
and standard representations of $S_3$ then give~\cite{Dong_2004}
\begin{align}
 \chi_0&=\frac{F_0+3G_0+2R_0}{6},&
 \chi_3&=\frac{F_0-3G_0+2R_0}{6},&
 \chi_2&=\frac{F_0-R_0}{3}\,,\notag\\
 \chi_{\frac{13}{5}}&=\frac{F_1+3G_1+2R_1}{6},&
 \chi_{\frac{8}{5}}&=\frac{F_1-3G_1+2R_1}{6},&
 \chi_{\frac{3}{5}}&=\frac{F_1-R_1}{3}\,.
 \label{eq:stablechars}
\end{align}
The last column counts one copy of the multiplicity space of the
two dimensional standard representation. Thus the $B$ vacuum
character is $F_0=\chi_0+\chi_3+2\chi_2$. The other ordinary
sectors give
\begin{align}
 \chi_{\frac{1}{2}}&=I_{\frac{1}{2}}T_0,&
 \chi_{\frac{3}{2}}&=I_0T_{\frac{3}{2}},&
 \chi_{\frac{11}{10}}&=I_{\frac{1}{2}}T_{\frac{3}{5}},&
 \chi_{\frac{1}{10}}&=I_0T_{\frac{1}{10}}\,.
 \label{eq:fourchars}
\end{align}
The two $B$ representations twisted by a 3-cycle have characters
\begin{align}
 H_0(\tau)&=\frac{\eta(\tau)}{\eta(\frac{\tau}{3})}
           \bigl(P_{\frac{1}{15}}-P_{\frac{2}{5}}-P_{\frac{7}{5}}\bigr)\,,\notag\\
 H_1(\tau)&=\frac{\eta(\tau)}{\eta(\frac{\tau}{3})}
           \bigl(P_0+P_3-P_{\frac{2}{3}}\bigr)\,.
 \label{eq:cyclechars}
\end{align}
The fractional levels in these representations are separated by
$\frac{1}{3}$. Resolving the three eigenvalues of the 3-cycle groups these
levels into three ordinary $D$ representations. For either of the two series
$H=\sum_e a_e q^e$, the projector is
\begin{align}
 [H]_h(\tau)
 =\frac13\sum_{j=0}^2
 e^{-2\pi i j(h-\frac{1}{20})}H(\tau+j)\,,
 \label{eq:cycleprojection}
\end{align}
where the retained powers obey $e-h+\frac{1}{20}\in\mathbb Z$.
Applying this projector gives
\begin{align}
 \chi_h&=[H_0]_h&&
 \left(h=\frac19,\frac79,\frac{13}{9}\right)\,,\notag\\
 \chi_h&=[H_1]_h&&
 \left(h=\frac{2}{45},\frac{17}{45},\frac{32}{45}\right)\,.
 \label{eq:cycleresolution}
\end{align}
Resolving the two eigenvalues of a transposition similarly gives
the eight characters
\begin{align}
 \chi_{\frac{1}{16}}&=I_{\frac{1}{16}}T_0,&
 \chi_{\frac{25}{16}}&=I_{\frac{1}{16}}T_{\frac{3}{2}}\,,\notag\\
 \chi_{\frac{13}{80}}&=I_{\frac{1}{16}}T_{\frac{1}{10}},&
 \chi_{\frac{53}{80}}&=I_{\frac{1}{16}}T_{\frac{3}{5}}\,,\notag\\
 \chi_{\frac{7}{16}}&=I_0T_{\frac{7}{16}},&
 \chi_{\frac{15}{16}}&=I_{\frac{1}{2}}T_{\frac{7}{16}}\,,\notag\\
 \chi_{\frac{3}{80}}&=I_0T_{\frac{3}{80}},&
 \chi_{\frac{43}{80}}&=I_{\frac{1}{2}}T_{\frac{3}{80}}\,.
 \label{eq:transchars}
\end{align}
There are therefore six characters from the two $S_3$ invariant
ordinary representations, four from the other ordinary orbits, six
from 3-cycles and eight from transpositions. Together they give
the twenty four $D$ representations~\cite{Dong_2004,Dong_2017}.
In particular, the vacuum expansion is
\begin{align}
 \chi_0(q)=q^{-\frac{1}{20}}
 \bigl(1+q^2+q^3+2q^4+2q^5+5q^6+5q^7+\cdots\bigr)\,.
 \label{eq:Dvacuumexpansion}
\end{align}

\bibliographystyle{JHEP}
\bibliography{references}

@article{Huang:2014ixa,
    author = "Huang, Yi-Zhi and Kirillov, Alexander and Lepowsky, James",
    title = "{Braided tensor categories and extensions of vertex operator algebras}",
    eprint = "1406.3420",
    archivePrefix = "arXiv",
    primaryClass = "math.QA",
    doi = "10.1007/s00220-015-2292-1",
    journal = "Commun. Math. Phys.",
    volume = "337",
    number = "3",
    pages = "1143--1159",
    year = "2015"
}

@article{MacAulay_1902, title={Some Formulae in Elimination}, volume={s1-35}, ISSN={0024-6115}, url={http://dx.doi.org/10.1112/plms/s1-35.1.3}, DOI={10.1112/plms/s1-35.1.3}, number={1}, journal={Proceedings of the London Mathematical Society}, publisher={Wiley}, author={MacAulay, F. S.}, year={1902}, month=may, pages={3--27} }

@article{Hohn:2023auw,
    author = {H{\"o}hn, Gerald and M{\"o}ller, Sven},
    title = "{Classification of Self-Dual Vertex Operator Superalgebras of Central Charge at Most 24}",
    eprint = "2303.17190",
    archivePrefix = "arXiv",
    primaryClass = "math.QA",
    month = "3",
    year = "2023"
}

@inbook{Lazard_1983, title={Gr{\"o}bner bases, Gaussian elimination and resolution of systems of algebraic equations}, ISBN={9783540387565}, ISSN={1611-3349}, url={http://dx.doi.org/10.1007/3-540-12868-9_99}, DOI={10.1007/3-540-12868-9_99}, booktitle={Computer Algebra}, publisher={Springer Berlin Heidelberg}, author={Lazard, D.}, year={1983}, pages={146--156} }

@article{Gaiotto:2012np,
    author = "Gaiotto, Davide",
    title = "{Domain Walls for Two-Dimensional Renormalization Group Flows}",
    eprint = "1201.0767",
    archivePrefix = "arXiv",
    primaryClass = "hep-th",
    doi = "10.1007/JHEP12(2012)103",
    journal = "JHEP",
    volume = "12",
    pages = "103",
    year = "2012"
}

@article{Fuchs:2002cm,
    author = "Fuchs, Jurgen and Runkel, Ingo and Schweigert, Christoph",
    title = "{TFT construction of RCFT correlators 1. Partition functions}",
    eprint = "hep-th/0204148",
    archivePrefix = "arXiv",
    reportNumber = "PAR-LPTHE-02-25",
    doi = "10.1016/S0550-3213(02)00744-7",
    journal = "Nucl. Phys. B",
    volume = "646",
    pages = "353--497",
    year = "2002"
}

@article{Dong_2004, title={{$\mathbb{Z}_3$} symmetry and {$W_3$} algebra in lattice vertex operator algebras}, volume={215}, ISSN={0030-8730}, url={http://dx.doi.org/10.2140/pjm.2004.215.245}, DOI={10.2140/pjm.2004.215.245}, number={2}, journal={Pacific Journal of Mathematics}, publisher={Mathematical Sciences Publishers}, author={Dong, Chongying and Lam, Ching Hung and Tanabe, Kenichiro and Yamada, Hiromichi and Yokoyama, Kazuhiro}, year={2004}, month=jun, pages={245--296} }

@article{Dong_2017, title={On orbifold theory}, volume={321}, ISSN={0001-8708}, url={http://dx.doi.org/10.1016/j.aim.2017.09.032}, DOI={10.1016/j.aim.2017.09.032}, journal={Advances in Mathematics}, publisher={Elsevier BV}, author={Dong, Chongying and Ren, Li and Xu, Feng}, year={2017}, month=dec, pages={1--30} }

@article{Ishikawa:2001zu,
    author = "Ishikawa, Hiroshi",
    title = "{Boundary states in coset conformal field theories}",
    eprint = "hep-th/0111230",
    archivePrefix = "arXiv",
    reportNumber = "TU-640",
    doi = "10.1016/S0550-3213(02)00131-1",
    journal = "Nucl. Phys. B",
    volume = "629",
    pages = "209--232",
    year = "2002"
}

@article{Recknagel:2002qq,
    author = "Recknagel, Andreas",
    title = "{Permutation branes}",
    eprint = "hep-th/0208119",
    archivePrefix = "arXiv",
    reportNumber = "KCL-MTH-01-45, ESI-1193-2002",
    doi = "10.1088/1126-6708/2003/04/041",
    journal = "JHEP",
    volume = "04",
    pages = "041",
    year = "2003"
}

@article{Fukusumi:2021zme,
    author = "Fukusumi, Yoshiki and Tachikawa, Yuji and Zheng, Yunqin",
    title = "{Fermionization and boundary states in 1+1 dimensions}",
    eprint = "2103.00746",
    archivePrefix = "arXiv",
    primaryClass = "hep-th",
    doi = "10.21468/SciPostPhys.11.4.082",
    journal = "SciPost Phys.",
    volume = "11",
    number = "4",
    pages = "082",
    year = "2021"
}

@article{Quella:2006de,
    author = "Quella, Thomas and Runkel, Ingo and Watts, Gerard M. T.",
    title = "{Reflection and transmission for conformal defects}",
    eprint = "hep-th/0611296",
    archivePrefix = "arXiv",
    reportNumber = "KCL-MTH-06-12, NSF-KITP-06-110",
    doi = "10.1088/1126-6708/2007/04/095",
    journal = "JHEP",
    volume = "04",
    pages = "095",
    year = "2007"
}

@article{Cardy:1989ir,
    author = "Cardy, John L.",
    title = "{Boundary Conditions, Fusion Rules and the Verlinde Formula}",
    reportNumber = "UCSB-TH-89-06",
    doi = "10.1016/0550-3213(89)90521-X",
    journal = "Nucl. Phys. B",
    volume = "324",
    pages = "581--596",
    year = "1989"
}

@article{Al-Ali:2016ndc,
    author = "Al-Ali, Masoumah and Linshaw, Andrew R.",
    title = "{The ${\mathbb{Z}_2}$ -Orbifold of the ${\mathcal{W}_3}$ -Algebra}",
    eprint = "1608.06255",
    archivePrefix = "arXiv",
    primaryClass = "math.RT",
    doi = "10.1007/s00220-016-2812-7",
    journal = "Commun. Math. Phys.",
    volume = "353",
    number = "3",
    pages = "1129--1150",
    year = "2017"
}

@article{Oshikawa:1996dj,
    author = "Oshikawa, Masaki and Affleck, Ian",
    title = "{Boundary conformal field theory approach to the critical two-dimensional Ising model with a defect line}",
    eprint = "cond-mat/9612187",
    archivePrefix = "arXiv",
    doi = "10.1016/S0550-3213(97)00219-8",
    journal = "Nucl. Phys. B",
    volume = "495",
    pages = "533--582",
    year = "1997"
}

@article{Bachas:2013ora,
    author = "Bachas, C. and Brunner, I. and Roggenkamp, D.",
    title = "{Fusion of Critical Defect Lines in the 2D Ising Model}",
    eprint = "1303.3616",
    archivePrefix = "arXiv",
    primaryClass = "cond-mat.stat-mech",
    reportNumber = "LMU-ASC-13-13, LPTENS-13-06",
    doi = "10.1088/1742-5468/2013/08/P08008",
    journal = "J. Stat. Mech.",
    volume = "1308",
    pages = "P08008",
    year = "2013"
}

@article{Kormos:2009sk,
    author = "Kormos, Marton and Runkel, Ingo and Watts, Gerard M. T.",
    title = "{Defect flows in minimal models}",
    eprint = "0907.1497",
    archivePrefix = "arXiv",
    primaryClass = "hep-th",
    reportNumber = "KCL-MTH-09-06",
    doi = "10.1088/1126-6708/2009/11/057",
    journal = "JHEP",
    volume = "11",
    pages = "057",
    year = "2009"
}

@incollection{DumasPernet2013,
  author    = {Dumas, Jean-Guillaume and Pernet, Cl{\'e}ment},
  title     = {Computational Linear Algebra over Finite Fields},
  booktitle = {Handbook of Finite Fields},
  editor    = {Mullen, Gary L. and Panario, Daniel},
  publisher = {CRC Press},
  year      = {2013},
  pages     = {514--528},
  url       = {https://arxiv.org/abs/1204.3735}
}

@article{Rong:2026xxg,
    author = "Rong, Junchen",
    title = "{A lattice construction of the Kormos-Runkel-Watts defect flows in minimal models}",
    eprint = "2611.03824",
    archivePrefix = "arXiv",
    primaryClass = "hep-th",
    month = "11",
    year = "2026"
}

@article{Chen:2026byx,
    author = "Chen, Jin",
    title = "{Renormalization Group Domain Walls from Phantom Symmetries}",
    eprint = "2610.23529",
    archivePrefix = "arXiv",
    primaryClass = "hep-th",
    reportNumber = "USTC-ICTS/PCFT-26-66",
    month = "10",
    year = "2026"
}

@article{Chen:2026nac,
    author = "Chen, Jin",
    title = "{When Conformal Interfaces Become Logarithmic}",
    eprint = "2610.23509",
    archivePrefix = "arXiv",
    primaryClass = "hep-th",
    reportNumber = "USTC-ICTS/PCFT-26-73",
    month = "10",
    year = "2026"
}

@article{Meineri:2025qwz,
    author = "Meineri, Marco and Radhakrishnan, Bharathkumar",
    title = "{The bootstrap of points and lines}",
    eprint = "2506.17382",
    archivePrefix = "arXiv",
    primaryClass = "hep-th",
    doi = "10.1007/JHEP04(2026)040",
    journal = "JHEP",
    volume = "04",
    pages = "040",
    year = "2026"
}

@article{Diatlyk:2023fwf,
    author = "Diatlyk, Oleksandr and Luo, Conghuan and Wang, Yifan and Weller, Quinten",
    title = "{Gauging non-invertible symmetries: topological interfaces and generalized orbifold groupoid in 2d QFT}",
    eprint = "2311.17044",
    archivePrefix = "arXiv",
    primaryClass = "hep-th",
    doi = "10.1007/JHEP03(2024)127",
    journal = "JHEP",
    volume = "03",
    pages = "127",
    year = "2024"
}

@article{Affleck:1991tk,
    author = "Affleck, Ian and Ludwig, Andreas W. W.",
    title = "{Universal noninteger 'ground state degeneracy' in critical quantum systems}",
    reportNumber = "UBCTP-91-007",
    doi = "10.1103/PhysRevLett.67.161",
    journal = "Phys. Rev. Lett.",
    volume = "67",
    pages = "161--164",
    year = "1991"
}

@article{Ishibashi:1988kg,
    author = "Ishibashi, Nobuyuki",
    title = "{The Boundary and Crosscap States in Conformal Field Theories}",
    reportNumber = "UT-530-TOKYO",
    doi = "10.1142/S0217732389000320",
    journal = "Mod. Phys. Lett. A",
    volume = "4",
    pages = "251",
    year = "1989"
}

@article{Behrend:1999bn,
    author = "Behrend, Roger E. and Pearce, Paul A. and Petkova, Valentina B. and Zuber, Jean-Bernard",
    title = "{Boundary conditions in rational conformal field theories}",
    eprint = "hep-th/9908036",
    archivePrefix = "arXiv",
    reportNumber = "SACLAY-SPH-T-99-085",
    doi = "10.1016/S0550-3213(99)00592-1",
    journal = "Nucl. Phys. B",
    volume = "570",
    pages = "525--589",
    year = "2000"
}

@article{Petkova:2000ip,
    author = "Petkova, V. B. and Zuber, J. B.",
    title = "{Generalized twisted partition functions}",
    eprint = "hep-th/0011021",
    archivePrefix = "arXiv",
    reportNumber = "UNN-SCM-M-00-07, CERN-TH-2000-322",
    doi = "10.1016/S0370-2693(01)00276-3",
    journal = "Phys. Lett. B",
    volume = "504",
    pages = "157--164",
    year = "2001"
}

@article{Friedan:2003yc,
    author = "Friedan, Daniel and Konechny, Anatoly",
    title = "{On the boundary entropy of one-dimensional quantum systems at low temperature}",
    eprint = "hep-th/0312197",
    archivePrefix = "arXiv",
    reportNumber = "RU-NHETC-2003-39",
    doi = "10.1103/PhysRevLett.93.030402",
    journal = "Phys. Rev. Lett.",
    volume = "93",
    pages = "030402",
    year = "2004"
}

@article{Verlinde:1988sn,
    author = "Verlinde, Erik P.",
    title = "{Fusion Rules and Modular Transformations in 2D Conformal Field Theory}",
    reportNumber = "THU-88/17",
    doi = "10.1016/0550-3213(88)90603-7",
    journal = "Nucl. Phys. B",
    volume = "300",
    pages = "360--376",
    year = "1988"
}

@inbook{Rocha_Caridi_1985, title={Vacuum Vector Representations of the Virasoro Algebra}, ISBN={9781461395508}, ISSN={0940-4740}, url={http://dx.doi.org/10.1007/978-1-4613-9550-8_22}, DOI={10.1007/978-1-4613-9550-8_22}, booktitle={Vertex Operators in Mathematics and Physics}, publisher={Springer US}, author={Rocha-Caridi, A.}, year={1985}, pages={451--473} }

@article{MathurMukhiSen1988,
  author  = {Mathur, Samir D. and Mukhi, Sunil and Sen, Ashoke},
  title   = {On the Classification of Rational Conformal Field Theories},
  journal = {Physics Letters B},
  volume  = {213},
  number  = {3},
  pages   = {303--308},
  year    = {1988},
  doi     = {10.1016/0370-2693(88)91765-0}
}

@article{MathurMukhiSen1989,
  author  = {Mathur, Samir D. and Mukhi, Sunil and Sen, Ashoke},
  title   = {Reconstruction of Conformal Field Theories from Modular Geometry on the Torus},
  journal = {Nuclear Physics B},
  volume  = {318},
  number  = {2},
  pages   = {483--540},
  year    = {1989},
  doi     = {10.1016/0550-3213(89)90615-9}
}

@article{NgRowellWangWen2023,
  author        = {Ng, Siu-Hung and Rowell, Eric C. and Wang, Zhenghan and Wen, Xiao-Gang},
  title         = {Reconstruction of Modular Data from {$SL_2(\mathbb{Z})$} Representations},
  journal       = {Communications in Mathematical Physics},
  volume        = {402},
  pages         = {2465--2545},
  year          = {2023},
  doi           = {10.1007/s00220-023-04775-w},
  eprint        = {2203.14829},
  archivePrefix = {arXiv},
  primaryClass  = {math.QA}
}

@misc{NgRowellWen2025,
  author        = {Ng, Siu-Hung and Rowell, Eric C. and Wen, Xiao-Gang},
  title         = {Classification of Modular Data up to Rank 12},
  year          = {2025},
  eprint        = {2308.09670v2},
  archivePrefix = {arXiv},
  primaryClass  = {math.QA},
  note          = {arXiv:2308.09670v2},
  url           = {https://arxiv.org/abs/2308.09670v2}
}

@article{BantayGannon2006,
  author        = {Bantay, Peter and Gannon, Terry},
  title         = {Conformal Characters and the Modular Representation},
  journal       = {Journal of High Energy Physics},
  volume        = {2006},
  number        = {02},
  pages         = {005},
  year          = {2006},
  doi           = {10.1088/1126-6708/2006/02/005},
  eprint        = {hep-th/0512011},
  archivePrefix = {arXiv}
}

@misc{SuBootstrap2026,
  author       = {Su, Ning},
  title        = {Bootstrapping {2d} Unitary Rational {CFT}},
  howpublished = {Talk at Bootstrap 2026, London},
  month        = aug,
  year         = {2026},
  url          = {https://indico.global/event/16114/contributions/162833/}
}

@article{Yin:2017yyn,
    author = "Yin, Xi",
    title = "{Aspects of Two-Dimensional Conformal Field Theories}",
    doi = "10.22323/1.305.0003",
    journal = "PoS",
    volume = "TASI2017",
    pages = "003",
    year = "2017"
}

@article{Zhang:2026gqp,
    author = "Zhang, Jin-Rui and Jin, Jing-Hao and Chen, Ting-Kai and Chen, Jin",
    title = "{Defect conformal manifolds along RG domain walls between {\ensuremath{\mathbb{Z}}}$_{N}$-parafermions and minimal models}",
    eprint = "2605.24978",
    archivePrefix = "arXiv",
    primaryClass = "hep-th",
    reportNumber = "USTC-ICTS/PCFT-26-31",
    doi = "10.1007/JHEP09(2026)123",
    journal = "JHEP",
    volume = "09",
    pages = "123",
    year = "2026"
}

@article{Chen:2026cqz,
    author = "Chen, Jin and Jia, Qiang",
    title = "{On general background of quantum non-invertible symmetry in 2D}",
    eprint = "2609.09550",
    archivePrefix = "arXiv",
    primaryClass = "hep-th",
    month = "9",
    year = "2026"
}

@article{Bachas:2001vj,
    author = "Bachas, C. and de Boer, J. and Dijkgraaf, R. and Ooguri, H.",
    title = "{Permeable conformal walls and holography}",
    eprint = "hep-th/0111210",
    archivePrefix = "arXiv",
    reportNumber = "CALT-68-2361, CITUSC-01-045, ITFA-2001-33, LPTENS-01-42",
    doi = "10.1088/1126-6708/2002/06/027",
    journal = "JHEP",
    volume = "06",
    pages = "027",
    year = "2002"
}

@article{Quella:2002ct,
    author = "Quella, Thomas and Schomerus, Volker",
    title = "{Symmetry breaking boundary states and defect lines}",
    eprint = "hep-th/0203161",
    archivePrefix = "arXiv",
    reportNumber = "AEI-2002-024, PAR-LPTHE-02-09",
    doi = "10.1088/1126-6708/2002/06/028",
    journal = "JHEP",
    volume = "06",
    pages = "028",
    year = "2002"
}

@article{Brunner:2007ur,
    author = "Brunner, Ilka and Roggenkamp, Daniel",
    title = "{Defects and bulk perturbations of boundary Landau-Ginzburg orbifolds}",
    eprint = "0712.0188",
    archivePrefix = "arXiv",
    primaryClass = "hep-th",
    doi = "10.1088/1126-6708/2008/04/001",
    journal = "JHEP",
    volume = "04",
    pages = "001",
    year = "2008"
}

@article{Cogburn:2023xzw,
    author = "Cogburn, Cameron V. and Fitzpatrick, Andrew Liam and Geng, Hao",
    title = "{CFT and lattice correlators near an RG domain wall between minimal models}",
    eprint = "2308.00737",
    archivePrefix = "arXiv",
    primaryClass = "hep-th",
    doi = "10.21468/SciPostPhysCore.7.2.021",
    journal = "SciPost Phys. Core",
    volume = "7",
    pages = "021",
    year = "2024"
}

@article{Antunes:2026kqb,
    author = "Antunes, Ant{\'o}nio and Rong, Junchen",
    title = "{On the Ising/tricritical-Ising interface: From ${W}_3$ symmetry to Rydberg atoms}",
    eprint = "2605.23848",
    archivePrefix = "arXiv",
    primaryClass = "hep-th",
    month = "5",
    year = "2026"
}

@article{Fukusumi:2024ejk,
    author = "Fukusumi, Yoshiki",
    title = "{Gauging or extending bulk and boundary conformal field theories: Application to bulk and domain wall problem in topological matter and their descriptions by mock modular covariant}",
    eprint = "2412.19577",
    archivePrefix = "arXiv",
    primaryClass = "hep-th",
    doi = "10.1103/f1bp-fzq5",
    journal = "Phys. Rev. B",
    volume = "112",
    number = "7",
    pages = "075144",
    year = "2025"
}

@article{Fukusumi:2026pkr,
    author = "Fukusumi, Yoshiki and Furuta, Yuma",
    title = "{Organizing transitions and their cascades: Generalized symmetry enforcement in massless flows or Higgs transitions}",
    eprint = "2608.08428",
    archivePrefix = "arXiv",
    primaryClass = "hep-th",
    reportNumber = "KYUSHU-HET-367",
    month = "8",
    year = "2026"
}

@article{Fukusumi:2026tii,
    author = "Fukusumi, Yoshiki and Nakashiba, Shuma",
    title = "{Difference between pair creation and anomaly matching: Nontrivial aspects of invertible symmetries and defects in $SU(N)$ models}",
    eprint = "2610.21478",
    archivePrefix = "arXiv",
    primaryClass = "hep-th",
    month = "10",
    year = "2026"
}

@article{Antinucci:2025uvj,
    author = "Antinucci, Andrea and Copetti, Christian and Galati, Giovanni and Rizi, Giovanni",
    title = "{Defect Conformal Manifolds from Phantom Noninvertible Symmetries}",
    eprint = "2505.09668",
    archivePrefix = "arXiv",
    primaryClass = "hep-th",
    doi = "10.1103/w43n-sybv",
    journal = "Phys. Rev. Lett.",
    volume = "135",
    number = "21",
    pages = "211602",
    year = "2025"
}

@article{Furuta:2025ahl,
    author = "Furuta, Yuma and Kusuki, Yuya and Onagi, Toshiki",
    title = "{Transmission coefficients from phantom currents}",
    eprint = "2511.00356",
    archivePrefix = "arXiv",
    primaryClass = "hep-th",
    reportNumber = "KYUSHU-HET-341, RIKEN-iTHEMS-Report-25",
    doi = "10.1103/fdfb-tkz4",
    journal = "Phys. Rev. D",
    volume = "113",
    number = "4",
    pages = "045008",
    year = "2026"
}

@article{Gaiotto:2014kfa,
    author = "Gaiotto, Davide and Kapustin, Anton and Seiberg, Nathan and Willett, Brian",
    title = "{Generalized Global Symmetries}",
    eprint = "1412.5148",
    archivePrefix = "arXiv",
    primaryClass = "hep-th",
    doi = "10.1007/JHEP02(2015)172",
    journal = "JHEP",
    volume = "02",
    pages = "172",
    year = "2015"
}

@article{Luo:2023ive,
    author = "Luo, Ran and Wang, Qing-Rui and Wang, Yi-Nan",
    title = "{Lecture notes on generalized symmetries and applications}",
    eprint = "2307.09215",
    archivePrefix = "arXiv",
    primaryClass = "hep-th",
    doi = "10.1016/j.physrep.2024.02.002",
    journal = "Phys. Rept.",
    volume = "1065",
    pages = "1--43",
    year = "2024"
}

@inproceedings{Shao:2023gho,
    author = "Shao, Shu-Heng",
    title = "{What's Done Cannot Be Undone: TASI Lectures on Non-Invertible Symmetries}",
    booktitle = "{Theoretical Advanced Study Institute in Elementary Particle Physics 2023}: {Aspects of Symmetry}",
    eprint = "2308.00747",
    archivePrefix = "arXiv",
    primaryClass = "hep-th",
    reportNumber = "YITP-SB-2023-19",
    month = "8",
    year = "2023"
}

@article{Bhardwaj:2017xup,
    author = "Bhardwaj, Lakshya and Tachikawa, Yuji",
    title = "{On finite symmetries and their gauging in two dimensions}",
    eprint = "1704.02330",
    archivePrefix = "arXiv",
    primaryClass = "hep-th",
    reportNumber = "IPMU-17-0049",
    doi = "10.1007/JHEP03(2018)189",
    journal = "JHEP",
    volume = "03",
    pages = "189",
    year = "2018"
}

@article{Chang:2018iay,
    author = "Chang, Chi-Ming and Lin, Ying-Hsuan and Shao, Shu-Heng and Wang, Yifan and Yin, Xi",
    title = "{Topological Defect Lines and Renormalization Group Flows in Two Dimensions}",
    eprint = "1802.04445",
    archivePrefix = "arXiv",
    primaryClass = "hep-th",
    reportNumber = "CALT-TH-2017-067, CALT-TH 2017-067, PUPT-2546",
    doi = "10.1007/JHEP01(2019)026",
    journal = "JHEP",
    volume = "01",
    pages = "026",
    year = "2019"
}

@article{Frohlich:2006ch,
    author = "Frohlich, Jurg and Fuchs, Jurgen and Runkel, Ingo and Schweigert, Christoph",
    title = "{Duality and defects in rational conformal field theory}",
    eprint = "hep-th/0607247",
    archivePrefix = "arXiv",
    reportNumber = "KCL-MTH-06-08, ZMP-HH-06-11",
    doi = "10.1016/j.nuclphysb.2006.11.017",
    journal = "Nucl. Phys. B",
    volume = "763",
    pages = "354--430",
    year = "2007"
}

@article{Gaiotto:2020iye,
    author = "Gaiotto, Davide and Kulp, Justin",
    title = "{Orbifold groupoids}",
    eprint = "2008.05960",
    archivePrefix = "arXiv",
    primaryClass = "hep-th",
    doi = "10.1007/JHEP02(2021)132",
    journal = "JHEP",
    volume = "02",
    pages = "132",
    year = "2021"
}

@article{Thorngren:2021yso,
    author = "Thorngren, Ryan and Wang, Yifan",
    title = "{Fusion category symmetry. Part II. Categoriosities at c = 1 and beyond}",
    eprint = "2106.12577",
    archivePrefix = "arXiv",
    primaryClass = "hep-th",
    doi = "10.1007/JHEP07(2024)051",
    journal = "JHEP",
    volume = "07",
    pages = "051",
    year = "2024"
}

@article{Nakayama:2024msv,
    author = "Nakayama, Yu and Tanaka, Takahilo",
    title = "{Infinitely many new renormalization group flows between Virasoro minimal models from non-invertible symmetries}",
    eprint = "2407.21353",
    archivePrefix = "arXiv",
    primaryClass = "hep-th",
    reportNumber = "YITP-24-92",
    doi = "10.1007/JHEP11(2024)137",
    journal = "JHEP",
    volume = "11",
    pages = "137",
    year = "2024"
}

@article{Ambrosino:2025yug,
    author = "Ambrosino, Federico and Negro, Stefano",
    title = "{Minimal Model Renormalization Group Flows: Noninvertible Symmetries and Nonperturbative Description}",
    eprint = "2501.07511",
    archivePrefix = "arXiv",
    primaryClass = "hep-th",
    reportNumber = "DESY-25-003",
    doi = "10.1103/dg1s-5vp6",
    journal = "Phys. Rev. Lett.",
    volume = "135",
    number = "2",
    pages = "021602",
    year = "2025"
}

@article{Gaberdiel:2026sfg,
    author = "Gaberdiel, Matthias R. and Merkens, Lasse",
    title = "{Defects in $ \mathcal{N}=1 $ minimal models and RG flows}",
    eprint = "2601.03879",
    archivePrefix = "arXiv",
    primaryClass = "hep-th",
    doi = "10.1007/JHEP05(2026)018",
    journal = "JHEP",
    volume = "05",
    pages = "018",
    year = "2026"
}

@article{Ambrosino:2026umb,
    author = "Ambrosino, Federico and Proch{\'a}zka, Tom{\'a}{\v{s}}",
    title = "{RG flows of minimal $\mathcal W$-algebra CFTs via non-invertible symmetries}",
    eprint = "2601.18667",
    archivePrefix = "arXiv",
    primaryClass = "hep-th",
    month = "1",
    year = "2026"
}

@article{Benedetti:2026drn,
    author = "Benedetti, Valentin and Fendley, Paul and Magan, Javier M.",
    title = "{Non-invertible symmetries and selection rules for RG flows of coset models}",
    eprint = "2603.09591",
    archivePrefix = "arXiv",
    primaryClass = "hep-th",
    month = "3",
    year = "2026"
}

@article{Fuchs:2001qc,
    author = "Fuchs, Jurgen and Schweigert, Christoph",
    title = "{Category theory for conformal boundary conditions}",
    eprint = "math/0106050",
    archivePrefix = "arXiv",
    reportNumber = "PAR-LPTHE-01-24",
    journal = "Fields Inst. Commun.",
    volume = "39",
    pages = "25",
    year = "2003"
}

@article{Frohlich:2004ef,
    author = "Frohlich, Jurg and Fuchs, Jurgen and Runkel, Ingo and Schweigert, Christoph",
    title = "{Kramers-Wannier duality from conformal defects}",
    eprint = "cond-mat/0404051",
    archivePrefix = "arXiv",
    reportNumber = "HU-EP-04-19",
    doi = "10.1103/PhysRevLett.93.070601",
    journal = "Phys. Rev. Lett.",
    volume = "93",
    pages = "070601",
    year = "2004"
}

@article{Fuchs:2012dt,
    author = "Fuchs, Jurgen and Schweigert, Christoph and Valentino, Alessandro",
    title = "{Bicategories for boundary conditions and for surface defects in 3-d TFT}",
    eprint = "1203.4568",
    archivePrefix = "arXiv",
    primaryClass = "hep-th",
    reportNumber = "HAMBURGER-BEITR-ZUR-MATHEMATIK-NR-433, ZMP-HH-12-5, HAMBURGER-BEITR.-ZUR-MATHEMATIK-NR.-433",
    doi = "10.1007/s00220-013-1723-0",
    journal = "Commun. Math. Phys.",
    volume = "321",
    pages = "543--575",
    year = "2013"
}

@article{Komargodski:2020mxz,
    author = "Komargodski, Zohar and Ohmori, Kantaro and Roumpedakis, Konstantinos and Seifnashri, Sahand",
    title = "{Symmetries and strings of adjoint QCD$_{2}$}",
    eprint = "2008.07567",
    archivePrefix = "arXiv",
    primaryClass = "hep-th",
    reportNumber = "YITP-SB-20-28",
    doi = "10.1007/JHEP03(2021)103",
    journal = "JHEP",
    volume = "03",
    pages = "103",
    year = "2021"
}

@article{Thorngren:2019iar,
    author = "Thorngren, Ryan and Wang, Yifan",
    title = "{Fusion category symmetry. Part I. Anomaly in-flow and gapped phases}",
    eprint = "1912.02817",
    archivePrefix = "arXiv",
    primaryClass = "hep-th",
    reportNumber = "PUPT-2603",
    doi = "10.1007/JHEP04(2024)132",
    journal = "JHEP",
    volume = "04",
    pages = "132",
    year = "2024"
}

@article{Kaidi:2022cpf,
    author = "Kaidi, Justin and Ohmori, Kantaro and Zheng, Yunqin",
    title = "{Symmetry TFTs for Non-invertible Defects}",
    eprint = "2209.11062",
    archivePrefix = "arXiv",
    primaryClass = "hep-th",
    doi = "10.1007/s00220-023-04859-7",
    journal = "Commun. Math. Phys.",
    volume = "404",
    number = "2",
    pages = "1021--1124",
    year = "2023"
}

@article{Chang:2022hud,
    author = "Chang, Chi-Ming and Chen, Jin and Xu, Fengjun",
    title = "{Topological defect lines in two dimensional fermionic CFTs}",
    eprint = "2208.02757",
    archivePrefix = "arXiv",
    primaryClass = "hep-th",
    doi = "10.21468/SciPostPhys.15.5.216",
    journal = "SciPost Phys.",
    volume = "15",
    number = "5",
    pages = "216",
    year = "2023"
}

@article{Kaidi:2023maf,
    author = "Kaidi, Justin and Nardoni, Emily and Zafrir, Gabi and Zheng, Yunqin",
    title = "{Symmetry TFTs and anomalies of non-invertible symmetries}",
    eprint = "2301.07112",
    archivePrefix = "arXiv",
    primaryClass = "hep-th",
    doi = "10.1007/JHEP10(2023)053",
    journal = "JHEP",
    volume = "10",
    pages = "053",
    year = "2023"
}

@article{Bhardwaj:2023ayw,
    author = "Bhardwaj, Lakshya and Schafer-Nameki, Sakura",
    title = "{Generalized charges, part II: Non-invertible symmetries and the symmetry TFT}",
    eprint = "2305.17159",
    archivePrefix = "arXiv",
    primaryClass = "hep-th",
    doi = "10.21468/SciPostPhys.19.4.098",
    journal = "SciPost Phys.",
    volume = "19",
    number = "4",
    pages = "098",
    year = "2025"
}

@article{Bhardwaj:2023kri,
    author = "Bhardwaj, Lakshya and Bottini, Lea E. and Fraser-Taliente, Ludovic and Gladden, Liam and Gould, Dewi S. W. and Platschorre, Arthur and Tillim, Hannah",
    title = "{Lectures on generalized symmetries}",
    eprint = "2307.07547",
    archivePrefix = "arXiv",
    primaryClass = "hep-th",
    doi = "10.1016/j.physrep.2023.11.002",
    journal = "Phys. Rept.",
    volume = "1051",
    pages = "1--87",
    year = "2024"
}

@article{Chen:2023qnv,
    author = "Chen, Jin and Cui, Wei and Haghighat, Babak and Wang, Yi-Nan",
    title = "{SymTFTs and duality defects from 6d SCFTs on 4-manifolds}",
    eprint = "2305.09734",
    archivePrefix = "arXiv",
    primaryClass = "hep-th",
    doi = "10.1007/JHEP11(2023)208",
    journal = "JHEP",
    volume = "11",
    pages = "208",
    year = "2023"
}

@article{Cordova:2024iti,
    author = "Cordova, Clay and Holfester, Nicholas and Ohmori, Kantaro",
    title = "{Representation theory of solitons}",
    eprint = "2408.11045",
    archivePrefix = "arXiv",
    primaryClass = "hep-th",
    doi = "10.1007/JHEP06(2025)001",
    journal = "JHEP",
    volume = "06",
    pages = "001",
    year = "2025"
}

@article{Chen:2025qub,
    author = "Chen, Jin and Duan, Zhihao and Jia, Qiang and Lee, Sungjay",
    title = "{Fermionic Non-invertible Symmetry Behind Supersymmetric ADE Solitons}",
    eprint = "2511.22129",
    archivePrefix = "arXiv",
    primaryClass = "hep-th",
    reportNumber = "USTC-ICTS/PCFT-25-54",
    month = "11",
    year = "2025"
}

@article{Antinucci:2025fjp,
    author = "Antinucci, Andrea and Copetti, Christian and Gai, Yuhan and Schafer-Nameki, Sakura",
    title = "{Categorical Anomaly Matching}",
    eprint = "2508.00982",
    archivePrefix = "arXiv",
    primaryClass = "hep-th",
    month = "8",
    year = "2025"
}

@article{Chen:2026wyn,
    author = "Chen, Jin and Duan, Zhihao and Jia, Qiang and Lee, Sungjay",
    title = "{Superstrip Algebras and Nonperturbative Spectra in Fermionic Theories}",
    eprint = "2609.12750",
    archivePrefix = "arXiv",
    primaryClass = "hep-th",
    reportNumber = "USTC-ICTS/PCFT-26-59, KIAS-P26044",
    month = "9",
    year = "2026"
}

@article{Douglas:2010ic,
    author = "Douglas, Michael R.",
    editor = "Das, Sumit R. and Shapere, Alfred D.",
    title = "{Spaces of Quantum Field Theories}",
    eprint = "1005.2779",
    archivePrefix = "arXiv",
    primaryClass = "hep-th",
    doi = "10.1088/1742-6596/462/1/012011",
    journal = "J. Phys. Conf. Ser.",
    volume = "462",
    number = "1",
    pages = "012011",
    year = "2013"
}

@misc{Poland2026,
  author        = {Poland, David},
  title         = {{virasoro\_blocks}: Software for the {2d} Bootstrap},
  year          = {2026},
  eprint        = {2610.39106},
  archivePrefix = {arXiv},
  primaryClass  = {hep-th},
  url           = {https://arxiv.org/abs/2610.39106}
}

@misc{Nakayama2026,
  author        = {Nakayama, Yu},
  title         = {What Crossing Symmetry Forbids Just Above {$c=1$}},
  year          = {2026},
  eprint        = {2611.03217},
  archivePrefix = {arXiv},
  primaryClass  = {hep-th},
  url           = {https://arxiv.org/abs/2611.03217}
}

@misc{KousvosPiazzaVichi2026,
  author        = {Kousvos, Stefanos R. and Piazza, Alessandro and Vichi, Alessandro},
  title         = {The {Virasoro} Bootstrap: From the {$c\sim 1$} Desert to Coupled {Potts}},
  year          = {2026},
  eprint        = {2611.03565},
  archivePrefix = {arXiv},
  primaryClass  = {hep-th},
  url           = {https://arxiv.org/abs/2611.03565}
}

@misc{BenjaminLanghoffLiThaler2026,
  author        = {Benjamin, Nathan and Langhoff, Kevin and Li, Wei and Thaler, Jesse},
  title         = {Constraining the Spectral Gap of {2d} {CFTs} Just Above {$c=1$}},
  year          = {2026},
  eprint        = {2611.07011},
  archivePrefix = {arXiv},
  primaryClass  = {hep-th},
  url           = {https://arxiv.org/abs/2611.07011}
}
\end{document}